\documentclass[prd,floatfix,showpacs,preprintnumbers,nofootinbib,superscriptaddress]{revtex4}
\usepackage{amsmath,amssymb}
\usepackage{color}
\usepackage{graphicx}

\newcommand\bef{\begin{figure}}
\newcommand\eef[1]{\label{fg:#1}\end{figure}}
\newcommand\beq{\begin{equation}}
\newcommand\eeq[1]{\label{#1}\end{equation}}
\newcommand\beqa{\begin{eqnarray}}
\newcommand\eeqa[1]{\label{#1}\end{eqnarray}}
\newcommand\bet{\begin{table}}
\newcommand\eet[1]{\label{tb:#1}\end{table}}
\newcommand\best{\begin{subtable}}
\newcommand\eest[1]{\label{stb:#1}\end{subtable}}
\newcommand\betb{\begin{center}\begin{tabular}}
\newcommand\eetb{\end{tabular}\end{center}}
\newcommand\beit{\begin{itemize}}
\newcommand\eeit{\end{itemize}}

\newcommand\fgn[1]{Figure \ref{fg:#1}}
\newcommand\eqn[1]{eq.\ (\ref{#1})}

\newcommand\tbn[1]{Table \ref{tb:#1}}

\begin{document}
\title{Spectroscopy of doubly-charmed baryons from lattice QCD}
\author{M.\ \surname{Padmanath}}
\email{padmanath@theory.tifr.res.in}
\affiliation{Institute of Physics, University of Graz, 8010 Graz, Austria.}

\author{Robert\ G.\ \surname{Edwards}}
\email{edwards@jlab.org}
\affiliation{Jefferson Laboratory, 12000 Jefferson Avenue,  Newport News, VA 23606, USA}

\author{Nilmani\ \surname{Mathur}}
\email{nilmani@theory.tifr.res.in}
\affiliation{Department of Theoretical Physics, Tata Institute of Fundamental
         Research,\\ Homi Bhabha Road, Mumbai 400005, India.}

\author{Michael\ \surname{Peardon}}
\email{mjp@maths.tcd.ie}
\affiliation{School of Mathematics, Trinity College, Dublin 2, Ireland}

\collaboration{For the Hadron Spectrum Collaboration}
%\author{\surname{Others}}
%\email{others@others.com}
%\affiliation{Others institute}

\begin{abstract}
We present the ground and excited state spectra of doubly charmed
baryons from lattice QCD with dynamical quark fields. 
Calculations are performed on
anisotropic lattices of size $16^3\times 128$, with inverse spacing in
temporal direction $a_{t}^{-1} = 5.67(4)$ GeV and with a pion mass of
about 390 MeV. A large set of baryonic operators that respect the
symmetries of the lattice yet which retain a memory of their continuum
analogues are used. These operators transform as irreducible
representations of SU(3)$_F$ symmetry for flavor, SU(4) symmetry for
Dirac spins of quarks and O(3) for spatial symmetry.
The distillation method is utilized to generate baryon correlation
functions which are analysed using the variational fitting method to extract
excited states. The lattice spectra obtained have baryonic
states with well-defined total spins up to $\frac{7}{2}$ and the pattern of 
low lying states does not support the diquark picture for doubly charmed 
baryons. On the contrary the calculated spectra
are remarkably similar to the expectations from models 
with an SU(6)$\times$ O(3) symmetry. Various spin dependent
energy splittings between the extracted states are also evaluated. 
\end{abstract}

\maketitle

%Section 1
\section{Introduction}
The study of hadrons containing charm quarks has recently undergone a
renaissance. This resurgence of interest started with the
discovery of new resonances in the charmonium system as well as a few charmed
baryons. So far, emphasis has been given to the study of the meson sectors
both theoretically and experimentally, while heavy baryon physics
has received substantially less attention. Similar to charm mesons, a
comprehensive study of charm baryons can provide similar insight into
the strong interaction. However, in comparison to the
many light and strange baryon states, only a handful of charm baryons
have been discovered and a reliable determination of the quantum
numbers of most of these observed charmed baryons has yet to be
made \cite{PDG}. Only very recently a few excited singly charmed
baryons were discovered. Similarly, there is no observation for
triply-charmed baryons although QCD clearly predicts such
states.  For doubly-charmed baryons, only SELEX
reported the discovery of five resonances and
interpreted those as $\Xi_{ccd}^{+}$(3443), $\Xi_{ccd}^{+}$(3520),
$\Xi_{ccu}^{++}$(3460), $\Xi_{ccu}^{++}$(3541) and
$\Xi_{ccu}^{++}(3780)$~\cite{Mattson:2002vu, Russ:2002bw}. Later they
confirmed the $\Xi_{ccd}^{+}$(3520) state in two different decay modes
($\Xi^+_{cc} \rightarrow \Lambda_c K^{-} \pi^{+}$; $\Xi^+_{cc}
\rightarrow p D^{+}K^{-}$) at a mass of $3518.7\pm 1.7$ MeV with an
average lifetime less than 33
$fs$~\cite{Ocherashvili:2004hi}. 
%Moreover, helicity angular
%distribution analysis also suggested that the states $ccd^{+}$(3443)
%and $ccu^{++}$(3460) and the states $\Xi_{ccd}^{+}$(3520) and
%$\Xi_{ccu}^{++}$(3541) form isospin doublets with isospin splittings
%17 and 21 MeV respectively. 
However, these states have {\it not} been
observed either by BABAR \cite{Aubert:2006qw}, Belle
\cite{Chistov:2006zj, Kato:2013ynr} in $e^{+}e^{-}$ annihilation experiments or by LHCb at baryon-baryon collider experiments at CERN~\cite{Aaij:2013voa}.  In
SELEX, production of doubly charm baryons with a large
cross section occurred through baryon-baryon interactions which is
totally inconsistent with fragmentation production.  Further, the
helicity angular distribution analysis also suggests that the pair of
states, \{$\Xi_{cc}^{+}$(3443), $\Xi_{cc}^{++}$(3460)\} and
\{$\Xi_{cc}^{+}$(3520), $\Xi_{cc}^{++}$(3541)\} form isospin doublets
with isospin splittings 17 and 21 MeV respectively. There is no
precise understanding of these unusually large isospin splittings
observed in the doubly charm baryons, unlike the small isospin
splittings in the case of light and singly charmed
baryons. An explanation may be possible if the coulombic
electro-magnetic effect is much larger than strong interaction effect, 
leading to these baryons having a very compact size~\cite{Brodsky:2011zs}. 
In summary, the experimental status of doubly-charmed baryons has not been 
settled, however it is expected that consolidated analysis of large
data collected from the ongoing experiments at LHCb and future experiments like
PANDA @FAIR, and Super Belle will shed further light.

Doubly-charmed baryons are interesting systems as they
provide a unique insight into the nature of the strong
force in the combined presence of both slowly moving heavy quarks
along with the relativistic motion of a light quark. 
The excited spectra of these states and the splittings between them
can help to understand how the collective degrees
of freedom give rise to excitations in these systems. A comparison of these excitations with the corresponding spectra of singly and triply-charmed baryons, where the number of charm quark is one less and one more respectively, will be helpful to get information about quark-quark interactions. 
Doubly-charmed baryons are
characterized by two widely separated scales: the low momentum scale,
of order $\Lambda_{QCD}$, of the light quark and the relatively heavy
charm quark mass. A doubly-heavy baryon can be treated as a bound
state of a heavy antiquark and a light quark in the limit when the
typical momentum transfer between the two heavy quarks is larger than
$\Lambda_{QCD}$~\cite{Savage:1990di, Savage:1990pr}. 
In this limit of quark-diquark symmetry,
$QQq \leftrightarrow \bar{Q}q$, one can get definite
prediction of spin dependent energy splittings~\cite{Brambilla:2005yk, Fleming:2005pd}. It was argued ~\cite{Bardeen:2003kt,Quigg:2011sb} that because of heavy-quark
symmetry, the doubly heavy baryons can be viewed as ultraheavy mesons, [$QQ$]$q$ $\sim Q'q$ and the hyperfine mass splittings in these systems are suppressed. Moreover, the doubly heavy baryon ground state of the form [$QQ$]$_{J=1}q$ will consist of chiral multiplets containing spin ($1/2^+,3/2^+$) heavy spin fields~\cite{Bardeen:2003kt}. It is thus interesting to study these spin splittings to 
determine if the charm quark is sufficiently heavy to respect this 
quark-diquark symmetry. 
Doubly-charmed baryons have been studied over the years
using various theoretical methods such as the
non-relativistic~\cite{Gershtein:2000nx,Kiselev:2002iy,Roberts:2007ni}
as well as relativistic ~\cite{Ebert:2002ig,Martynenko:2007je} quark
models, heavy-quark effective theory~\cite{Korner:1994nh}, QCD sum
rules~\cite{Bagan:1992za,Kiselev:2001fw,Zhang:2008rt,Narison:2010py,Wang:2010hs}, Feynman-Hellmann theorem~\cite{Roncaglia:1995az}, mass formula~\cite{Burakovsky:1997vm} and Skyrmion model~\cite{Rho:1990uy}.

In light of existing and future experimental efforts to observe
doubly-charmed baryons, it is desirable to have first principle 
predictions from lattice QCD.  A quantitative
description of the spectra of doubly-charmed baryons from the
non-perturbative method of lattice QCD is valuable as it will enable
a comparison between the lattice-computed spectra of doubly-charmed
baryons to those obtained from potential models which have been 
successful for charmonia.  Moreover, all results from such a first
principles calculation will be predictions and thus naturally can
provide crucial inputs to current and future experimental discovery.
Given this significance of doubly-charmed baryons, it is 
desirable to study these states comprehensively using 
lattice QCD. Lattice QCD groups have studied the ground states of the 
doubly-charmed baryons using quenched
NRQCD~\cite{Mathur:2002ce}, quenched QCD with relativistic
quarks~\cite{Lewis:2001iz,Flynn:2003vz,Chiu:2005zc}, and full
QCD~\cite{Na:2007pv,Liu:2009jc,Lin:2010wb,Briceno:2012wt,Alexandrou:2012xk,Basak:2012py,Bali:2012ua,Namekawa:2013vu}
calculations.
However, all previous lattice calculations involve only the spin 1/2
and spin 3/2 ground state spectrum of $\Xi_{cc}$ and
$\Omega_{cc}$. It is expected that much more
information about the interactions between two charm quarks and
between charm and light quarks can be obtained by computing the
excited state spectra of these baryons, including in particular the
spin-dependent energy splittings, as well as by studying similar
spectra for other spin-parity channels.  Towards this goal, 
we report here the first attempt to compute the excited state spectra
of doubly-charmed baryons using dynamical lattice QCD. The ground
states for each spin-parity channel as well as their excited states up
to spin 7/2 are computed and a few spin-dependent energy splittings
are also studied. Similar studies of triply-charmed baryons have already
been reported in Ref.~\cite{Padmanath:2013zfa}, and in a subsequent
publication we will report results on singly charmed baryons.

To extract the excited states of doubly-charmed baryons we follow the
same procedure used in previous calculations for mesons
~\cite{Dudek:2007wv,Dudek:2010wm, Dudek:2009qf, Dudek:2010ew,Liu:2012ze,Moir:2013ub} and baryons~\cite{Edwards:2011jj,Dudek:2012ag,Edwards:2012fx,Padmanath:2013zfa}.  The gauge
configurations we use are generated with 2+1 flavour clover
fermions on anisotropic lattices~\cite{Edwards:2008ja,Lin:2008pr}. For
the charm quarks we also use clover fermions which are ${\cal{O}}(a)$
improved at tree level in tadpole-improved perturbation theory. A
large set of baryon operators which are first constructed in the
continuum and then subduced into various lattice
irreps~\cite{Edwards:2011jj} are used for this calculation. These
operators transform as irreducible representations of $SU(3)_F$
symmetry for flavor, $SU(4)$ symmetry for Dirac spins of quarks and
$O(3)$ symmetry for orbital angular momenta. Baryon correlation
functions are generated by using the ``distillation
method''~\cite{Peardon:2009gh} and then variational fitting method is utilized
to extract excited energies as well as to reliably determine the spins
of these states.

The layout of the paper is as follows.  In the next
section, we briefly describe lattice details, operator construction, 
the construction and analysis of correlation functions and
our procedure for identifying the continuum spins of extracted
states, which are already detailed earlier in
Refs.~\cite{Dudek:2007wv,Dudek:2010wm, Dudek:2009qf, Dudek:2010ew,Liu:2012ze,Moir:2013ub,Peardon:2009gh,Edwards:2011jj,Dudek:2012ag,Edwards:2012fx,Padmanath:2013zfa}.
In section III we present our results giving details of energy
splittings in subsection IIIA.  Finally, a summary of the work is
presented in section IV.

\section{Computational Methods}
Over the last several years the Hadron Spectrum Collaboration (HSC) has
adopted a dynamical anisotropic lattice formulation to extract highly
excited hadron spectra. In this approach one uses a much finer
temporal lattice spacing than in the spatial directions and 
exploits this higher resolution to extract highly excited states which decay
rapidly at large Euclidean time separations with increasing noise
to signal ratio. On the other hand, this avoids the computational
cost that would come if one reduces the spacing in all directions.

\subsection{The lattice action}
We used the tree-level Symanzik-improved gauge action and the anisotropic
Shekholeslami-Wohlert fermion action with tree-level tadpole
improvement and three-dimensional stout-link smearing of gauge
fields for this work. The details of the formulation of actions 
as well as the method used to tune the anisotropic parameters can be 
found in Refs.~\cite{Edwards:2008ja, Lin:2008pr}.
\bet[h]
\centering
\betb{cccc|ccc}
\hline
Lattice size & $a_t m_\ell$   & $a_t m_s$ & $N_{\mathrm{cfgs}}$ & $m_\pi/$MeV & $m_K/m_{\pi}$ &$a_t m_\Omega$ \\\hline
$16^3 \times 128$ & $-0.0840$ & $-0.0743$ & 96 & 391&1.39        & $0.2951(22)$      
\\\hline
\eetb
\caption{Details of the gauge-field ensembles used. $N_{\mathrm{cfgs}}$  is the number of gauge-field configurations.}
\eet{lattices}
In \tbn{lattices} we show the lattice action parameters of the
gauge-field ensembles used in this work. We used the $\Omega$-baryon
mass to determine the lattice spacing and obtained $a_t^{-1} =
5.67(4)$ GeV. With an anisotropy of close to 3.5, this leads to $a_s
=$ 0.12 fm, and total volume $V \sim (1.9 fm)^3$.  We assume this volume is 
enough for a first study of doubly-charmed baryons.

The charm quark action used here is similar to the light quark sector
and the details are given in Ref.~\cite{Liu:2012ze}. The $\eta_c$
meson ground-state was used to determine the bare charm quark mass. By studying the
dispersion relation at low momenta, the action is made 
relativistic.  As mentioned in~\cite{Liu:2012ze}, it is expected that
the effects due to the absence of dynamical charm quark fields in this
calculation will be small.

\subsection{Baryon operators}
Following Ref.\cite{Edwards:2011jj} we construct a large set of baryon
operators for doubly-charmed baryons. In summary the construction has two 
steps: a set of continuum operators with well-defined
continuum spin is found and then these operators are subduced to the
irreducible representations (irreps) of the octahedral group on the lattice. A
set of continuum baryon interpolating operators with well-defined
continuum spin are constructed as 
\beq O^{[J^P]} \sim \left[
  \mathcal{F}_{\Sigma_F}\otimes\mathcal{S}_{\Sigma_S}\otimes\mathcal{D}_{\Sigma_D}
  \right] ^{J^P} \,,
\eeq{genericbaryon} 
where $\cal{F}$, $\cal{S}$ and
$\cal{D}$ represent flavor, Dirac spin and spatial structure
respectively while the subscripts $\Sigma_i$ specify the permutation
symmetry in the respective subspaces. The details of these permutation
symmetries and their combinations are given in
Refs.\cite{Edwards:2011jj, Edwards:2012fx}.

By permutation symmetry, one can argue that the flavor structure of
$QQq$ and $QQs$ will be the same as $qqs$ combinations, which are the
flavor structures of light $\Sigma$ baryons. Hence the possible 
flavor-symmetry structures for doubly charm baryons are totally symmetric (S)
in all three flavor labels belonging to the decuplet flavor
constructions ($10_F$) as well as symmetric (MS) and antisymmetric
(MA) in the first two flavor labels belonging to the octet flavor
constructions ($8_F$).  For the decuplet structure, the flavor labels
being symmetric, the remaining spin and spatial part should be
combined symmetrically to form an overall symmetric interpolating operator. For
the octet structure, the flavor structure being MS and MA, the symmetry in
the remaining part of the baryon operator, excluding the color labels,
should be MS and MA to form a symmetric interpolating operator.

The spatial- as well as spin-symmetry combinations used are also presented 
in Ref.~\cite{Edwards:2011jj}. Up to two covariant derivatives are
considered and combined so as to transform as
orbital angular momentum, $L$, with maximum accessible values of 0, 1
and 2.  The subset of operators formed by considering only the 
upper two-component of the four-component Dirac-spinor in the Dirac-Pauli 
representation are labelled non-relativistic, while all other operators are
called relativistic. Another subset of operators with $D = 2$ and $L = 1$  in the
mixed symmetric and mixed antisymmetric combinations are identified as
hybrid operators because of their essential gluonic content; these operators 
vanish in the absence of a gluon field. 
In \tbn{n_operators}, we
show the allowed spin-parity patterns based on non-relativistic
quark spinors for up to two covariant derivatives.  Note
that with the non-relativistic operators, it is not possible to
construct a negative-parity state beyond spin $\frac52^-$, even with
non-local operators using two derivatives.  Use of relativistic
operators along with non-relativistic ones enable us to extract higher
negative-parity states as well as higher excited states.
\bet[h]	
\centering
\betb{| cc cc  cc  cc | lc lc lc lc| } \hline
D&&SU(3)$_F$&&    S    &&L&&    \multicolumn{8}{c|}{$J^P$}                          \\ \hline
0&& $8_F$   &&$\frac12$&&0&&$\frac12^+$ &&            &&            &&            & \\ 
 && $10_F$  &&$\frac32$&&0&&            &&$\frac32^+$ &&            &&            & \\ \hline
 &&         &&   $N_0$ && &&$\mathbf{1}$&&$\mathbf{1}$&&$\mathbf{0}$&&$\mathbf{0}$& \\ \hline
 && $8_F$   &&$\frac12$&&1&&$\frac12^-$ &&$\frac32^-$ &&            &&            & \\ 
1 &&         &&$\frac32$&&1&&$\frac12^-$ &&$\frac32^-$ &&$\frac52^-$ &&            & \\ 
 && $10_F$  &&$\frac12$&&1&&$\frac12^-$ &&$\frac32^-$ &&            &&            & \\ \hline
 &&         &&    $N_1$&& &&$\mathbf{3}$&&$\mathbf{3}$&&$\mathbf{1}$&&$\mathbf{0}$& \\ \hline
 && $8_F$   &&$\frac12$&&0&&$\frac12^+$ &&            &&            &&            & \\
 &&         &&$\frac12$&&0&&$\frac12^+$ &&            &&            &&            & \\
 &&         &&$\frac12$&&1&&$\frac12^+$ &&$\frac32^+$ &&            &&            & \\
 &&         &&$\frac12$&&2&&            &&$\frac32^+$ &&$\frac52^+$ &&            & \\
 &&         &&$\frac12$&&2&&            &&$\frac32^+$ &&$\frac52^+$ &&            & \\
2 &&         &&$\frac32$&&0&&            &&$\frac32^+$ &&            &&            & \\ 
 &&         &&$\frac32$&&2&&$\frac12^+$ &&$\frac32^+$ &&$\frac52^+$ &&$\frac72^+$ & \\ 
 && $10_F$  &&$\frac12$&&0&&$\frac12^+$ &&            &&            &&            & \\
 &&         &&$\frac12$&&2&&            &&$\frac32^+$ &&$\frac52^+$ &&            & \\
 &&         &&$\frac32$&&0&&            &&$\frac32^+$ &&            &&            & \\ 
 &&         &&$\frac32$&&2&&$\frac12^+$ &&$\frac32^+$ &&$\frac52^+$ &&$\frac72^+$ & \\ \hline
 &&         && $N_2$   && &&$\mathbf{6}$&&$\mathbf{8}$&&$\mathbf{5}$&&$\mathbf{2}$& \\ \hline \hline
 && $8_F$   &&$\frac12$  &&1&&$\frac12^+$ &&$\frac32^+$ &&            &&            & \\
2 &&         &&$\frac32$  &&1&&$\frac12^+$ &&$\frac32^+$ &&$\frac52^+$ &&            & \\
 && $10_F$  &&$\frac12$  &&1&&$\frac12^+$ &&$\frac32^+$ &&            &&            & \\\hline
 &&         &&$N_{2}(hy)$&& &&$\mathbf{3}$&&$\mathbf{3}$&&$\mathbf{1}$&&$\mathbf{0}$&  \\ \hline
\eetb
\caption{Allowed spin-parity patterns based on non-relativistic quark
  spinors.  For each covariant derivative (D = 0, 1 and 2), the SU(3) flavor, 
  the quark
  spin $S$ and orbital angular momentum $L$ are listed followed by the
  allowed $J^{P}$ values. The total number of operators is listed as
  $N_i$ for derivative $i$. The lower part of the table gives the 
  two-derivative hybrid operators based on non-relativistic quark
  spinors, as discussed in~\cite{Dudek:2012ag}.}
\eet{n_operators}
These continuum operators are then subduced to the irreps of the cubic
group. The three irreps of the double-valued representations of the
octahedral group for half-integer spins are $G_1$, $G_2$ and $H$. The
details of this subduction procedure to obtain the lattice operators
was discussed in Ref.~\cite{Edwards:2011jj}. In \tbn{operators} we show
the number of operators that we obtained after subduction and are used
for this study. Both the number of positive ($g$), negative ($u$)
parity operators as well as the number of operators with
non-relativistic quark spinors (NR) and with hybrid content are shown
in this table.
\bet[h]	
\centering
\betb{ c | c  c  c  c  c  c }
\hline \hline
       &\multicolumn{2}{c}{$G_1$}&\multicolumn{2}{c}{H}&\multicolumn{2}{c}{$G_2$} \\ \cline{2-7}
       &   $g$    &     $u$      &    $g$   &    $u$   &     $g$   &    $u$       \\ \hline
Total  &    55    &      55      &     90   &     90   &      35   &     35       \\
Hybrid &    12    &      12      &     16   &     16   &       4   &      4       \\
NR     &    11    &       3      &     19   &      4   &       8   &      1       \\
\hline
\eetb
\caption{The total number of operators, with covariant derivatives up to two,  obtained after subduction to various irreps. The number of
  non-relativistic (NR) and hybrid operators for each irreps and for
  both parities are also mentioned.}
\eet{operators}
%===========================================================
\subsection{Analysis of baryon correlators : Distillation and variational methods}
After constructing a large set of operators for each
irrep $\Lambda$, we calculate the matrix of correlation functions
\beq C_{ij}(t\equiv t_f-t_i) = \langle 0
|O_i(t_f)O_{j}^{\dagger}(t_i)|0\rangle.  \eeq{2pt} between a baryon source at
time $t_i$ and sink at time $t_f$. 
The {\it distillation method} \cite{Peardon:2009gh} provides an efficient 
means of constructing the large correlation matrices needed for this analysis. 
For this work, the method was realized by constructing the 
distillation operator with 64 eigenvectors of the gauge-covariant Laplacian 
and then correlation functions are computed from 4 time sources.
The matrix of correlation functions was then analyzed using
 the variational method, which proceeds by solving a generalized eigenvalue 
problem  of the form 
\beq C_{ij}(t)v_{j}^{n} =
\lambda_{n}(t,t_0)C_{ij}(t_0)v_{j}^{n}, 
\eeq 
where $C(t)$ is the
matrix of correlators at time-slice $t$, i.e.  $C_{ij}(t) =
\langle\mathcal{O}_i(t)\mathcal{O}_j(0)\rangle$, $\lambda_{n}(t,t_0)$
are the {\it principal correlators} and $v_{j}^{n}$'s are the
eigenvectors which determine the overlap of an operator to a particular state. 
The details
of our fitting procedure is given in Ref.~\cite{Dudek:2007wv} and was utilized
 to extract the excited state spectra of mesons~\cite{Dudek:2010wm, Dudek:2009qf, Dudek:2010ew,Liu:2012ze,
  Moir:2013ub} as well as of baryons~\cite{Edwards:2011jj, Dudek:2012ag,Edwards:2012fx,Padmanath:2013zfa}. In \fgn{prin_corr_ccs_p} and \fgn{prin_corr_ccu_p}, we
plot some examples of fits to the principal correlators for $\Omega_{cc}$ baryons in the $H_g$ irrep and for $\Xi_{cc}$ baryon in the $G_{2g}$ irrep.
\bef[tbh]
\centering
\hspace*{-0.25in}
\includegraphics[scale=1.3]{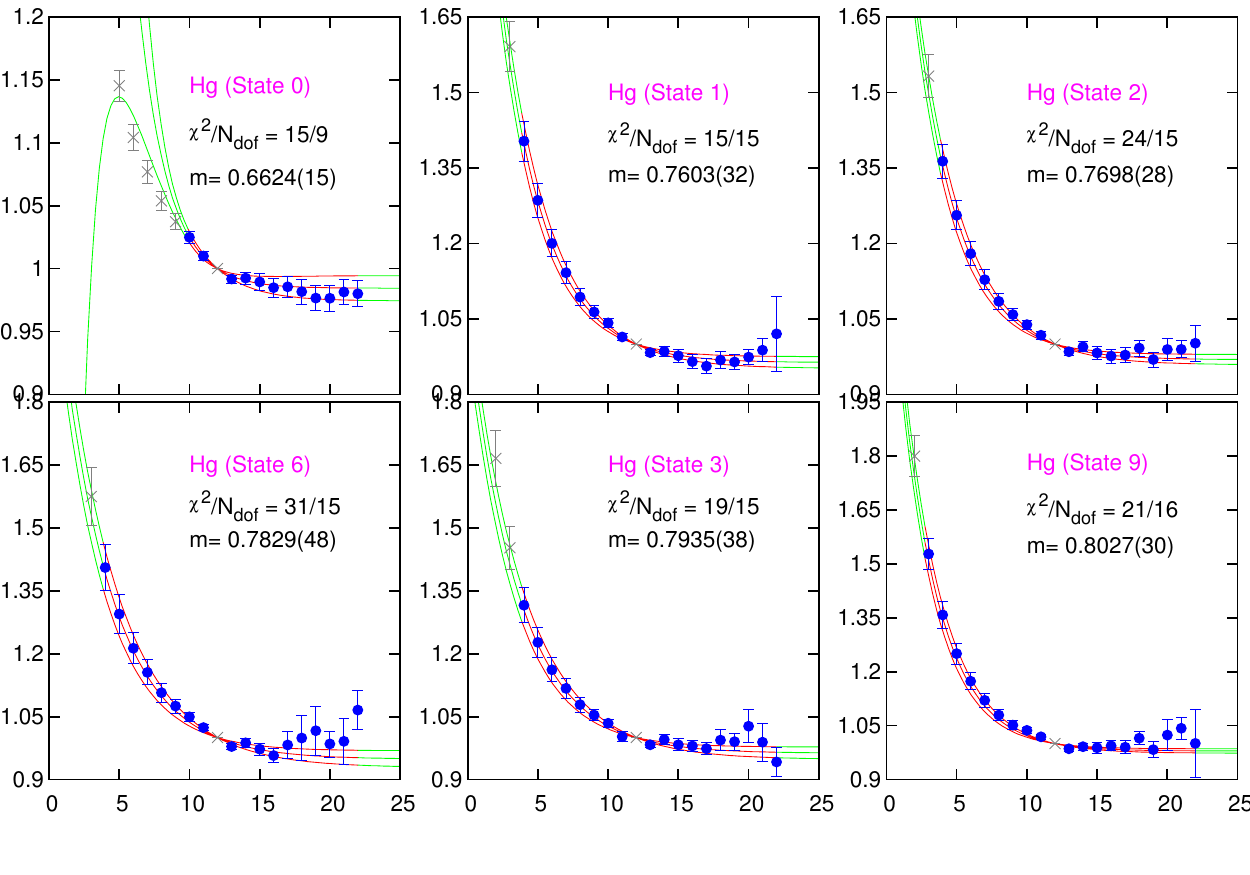}
\caption{Principal correlator fits for six states of $\Omega_{cc}$ baryons in irrep. $H_{g}$ that are identified as $J = 3/2^+$. Data points are obtained from $e^{m_n(t-t_0)}\lambda_n(t)$. Fits are carried out using a fitting form
 $\lambda_n(t) = (1-A_n)e^{-m_n(t-t_0)} +A_n e^{-m'_n(t-t_0)}$ with three fit parameters, $m_n$ : the state that we are looking for, $m'_n$ : where all other excited states are grouped together, and $A_n$ : which is related to overlap factor. 
The lines show the fits and one sigma-deviation according with $t_0 = 12$; the grey points are not included in the fits.}
\eef{prin_corr_ccs_p}
 
\bef[tbh]
\centering
\hspace*{-0.25in}
\includegraphics[scale=1.3]{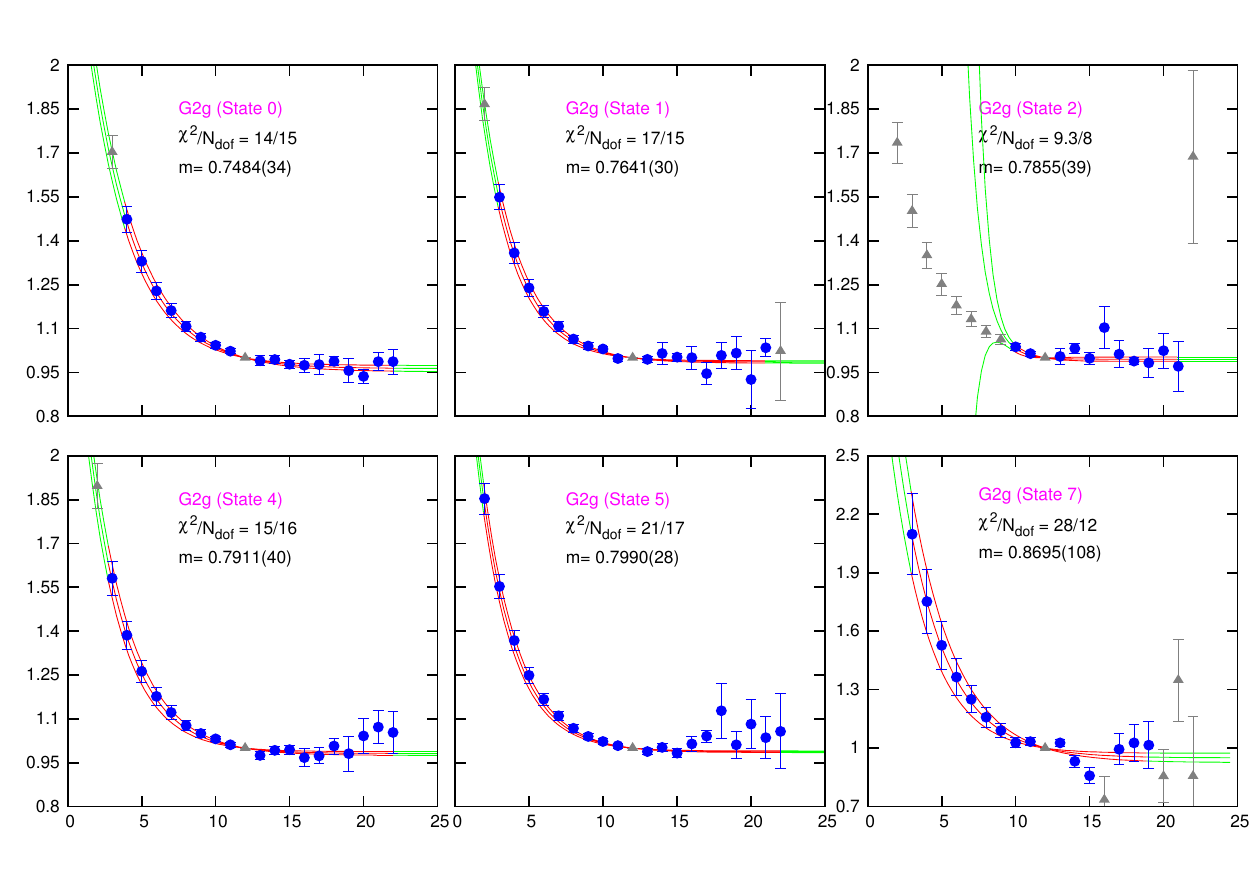}
\caption{Same as \fgn{prin_corr_ccu_p}, but for $\Xi_{cc}$ baryons in irrep. $G_{2g}$.}
\eef{prin_corr_ccu_p}

\subsection{Rotational symmetry}

We constructed operators in the continuum which were then subduced onto
the lattice irreps to form lattice operators. This allows us to determine the 
properties of these states in the continuum with some confidence.  Following 
Refs.~\cite{Edwards:2011jj, Dudek:2012ag,Edwards:2012fx,Padmanath:2013zfa}, 
\fgn{rel_ccu_Hg_corr_matplot} shows the normalized correlation functions, 
$C_{ij}/\sqrt{C_{ii}C_{jj}}$, for $\Xi_{cc}$ baryons transforming under 
$H_g$ and determined at time-slice 5.  The
normalization ensure diagonal entries are unity and off-diagonal entries 
are less than 1. Various operators are
represented by following abbreviations : non-relativistic (n),
relativistic (r), non-hybrid (1) and hybrid (2).  There are 90
operators used in this irrep, including operators up to two
derivatives. The solid lines divide these operators into spins
$\frac32, \frac52$ and $\frac72$, and the dashed lines separates operators 
defined above.  As is evident from the figure, the matrix is close to block 
diagonal implying that there are small correlations between 
operators subduced from different continuum spins. This suggests that there 
remains a remarkable degree of rotational symmetry in the matrices of 
correlation functions obtained for these lattice operators, similar to the 
light, strange and triply-charmed
baryons~\cite{Edwards:2012fx,Padmanath:2013zfa}.  Similar
block-diagonal matrices of correlation functions are observed in
other irreps and for $\Omega_{cc}$ baryons as well. 
%For the
%unambiguous identification of the continuum spin of states, these
%approximately orthogonal matrices are a necessary ingredient.

From the correlators one can also identify the
flavour mixing between different operators. For example, it is evident
from \fgn{rel_ccu_Hg_corr_matplot} that there is strong mixing between
relativistic operators belonging to octet and decuplet. However, this
flavour mixing is less evident for non-relativistic operators. As was
observed %in Ref. \cite{Meinel:2012qz}, 
in our previous study on 
triply-charmed baryons~\cite{Padmanath:2013zfa}, we found 
that there are
additional suppression in mixing for non-relativistic operators 
with a given $J$, but
with different $L$ and $S$, compared to those with the same $J$, as
well as the same $L$ and $S$. However, this suppression is not
present for relativistic operators. On the contrary, in most of the
cases involving relativistic operators, we found that mixings are comparable
for operators with given $J$, but with different $L$ and $S$ to
those of operators with the same $J$, as well as same $L$ and $S$.

\bef[tbh]
\includegraphics[scale=0.3]{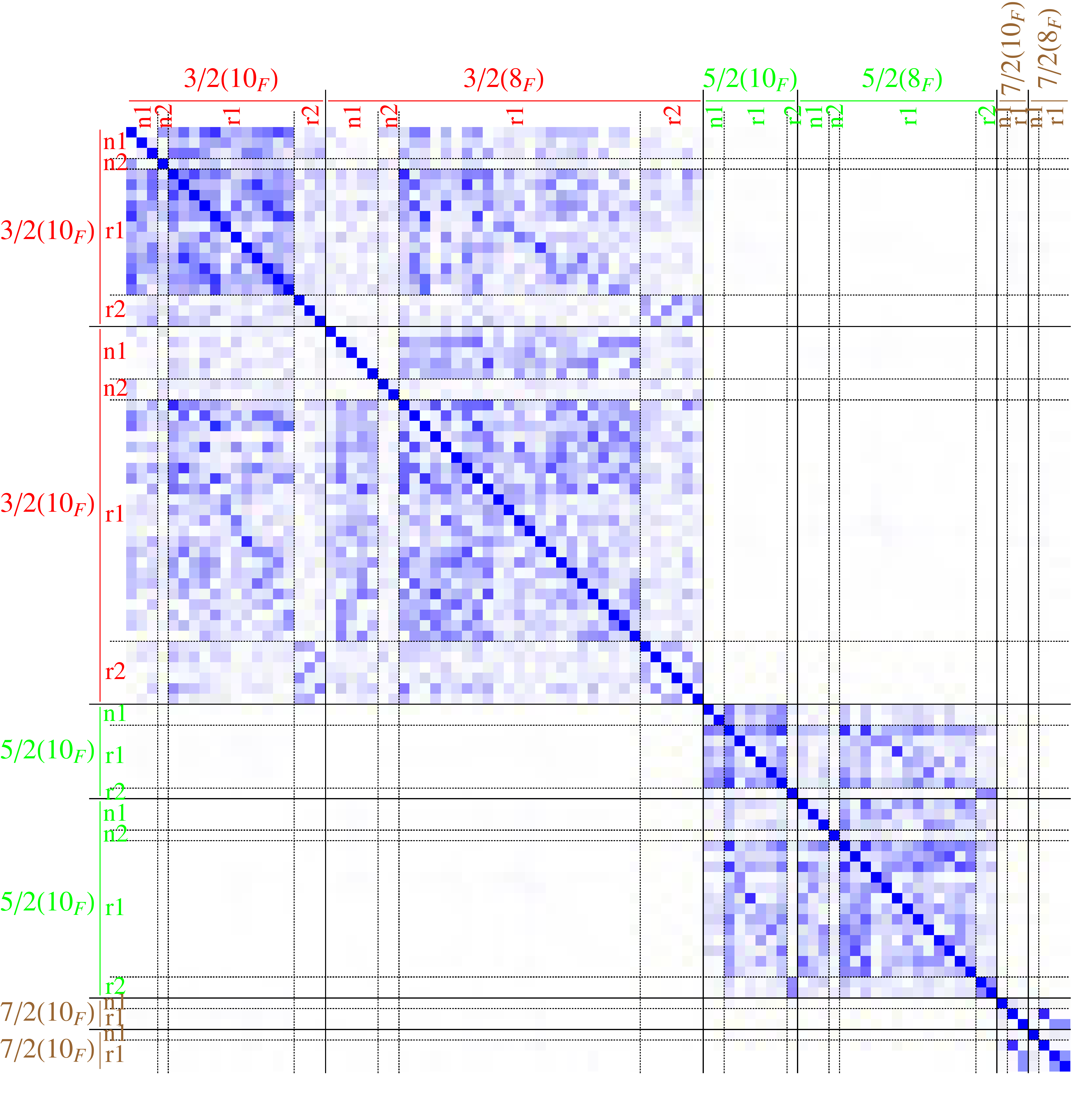}
\includegraphics[scale=0.1]{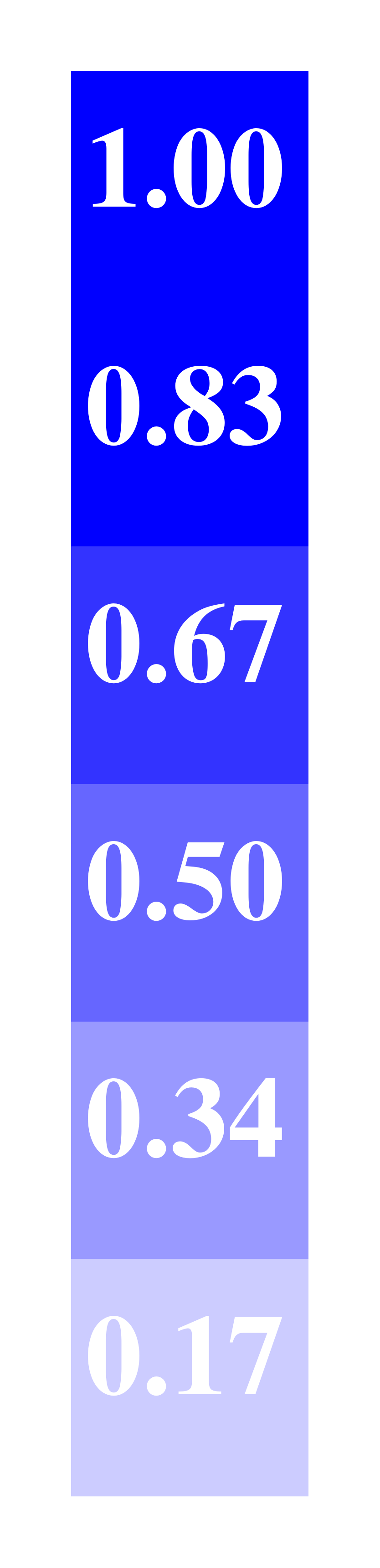}
\caption{The normalized correlation matrix, $C_{ij}/\sqrt{C_{ii}C_{jj}}$, at $t/a_t = 5$ are shown for $\Xi_{cc}$ baryons in the $H_g$ irrep.,
according to the darkness scale at the side. 
The operators are ordered such that those subduced from spin  3/2 appear first followed by spin 5/2 and then spin 7/2. The correlation matrix is observed to be mostly block diagonal in terms of spins which signifies the rotational symmetry on our lattice. There is significant mixing between octet ($8_F$) and decuplet ($10_F$) operators which is not present in the corresponding plot for light quark $\Sigma(uus)$ having same operator structures (see figure 2 of Ref.\cite{Edwards:2012fx}). This signifies that the $SU(3)$ flavor symmetry is badly broken for doubly-charmed $\Xi_{cc}$ baryons. Non-relativistic (n), relativistic (r),
non-hybrid (1) and hybrid (2) operators are also identified for each of those spins. The strength of flavor mixing is maximum between relativistic operators (r), while for non-relativistic operators this mixing is relatively mild.}
\eef{rel_ccu_Hg_corr_matplot}
%\bef[tbh]
%\includegraphics[scale=0.3]{fig/ccs_corr_matrix_Hg_t5.pdf}
%\includegraphics[scale=0.1]{fig/legend_matrix_plot.pdf}
%\caption{Same as \fgn{rel_ccu_Hg_z}, but for $\Omega_{cc}$ baryons in the irrep $H_g$. Here also one can notice significant flavor mixing between octet and decuplet states, in particular, between relativistic operators.}
%\eef{rel_ccs_Hg_corr_matplot}

\subsubsection{Continuum spin identification}
One main difficulty of lattice calculations of spectra is the identification of 
the spin of an extracted state at finite lattice spacing, particularly for the 
state onto which various operators from different irreps contribute. In 
Ref. ~\cite{Dudek:2007wv} a well defined procedure is adopted for spin 
identification by using the overlap factor. The {\it overlap-factors} of an operator $O_i$ to a state $n$ are $Z^n_i \equiv \langle n|O_i^{\dagger}|0\rangle$ 
which can be shown~\cite{Dudek:2007wv} to occur in the spectral decomposition 
of the matrices of the correlation functions as,
\beq{corr}
C_{ij}(t) = \sum_{n} {Z^{n*}_iZ^n_j\over 2 m_n} e^{-m_nt}.
\eeq{corr} 
One can use the orthogonality for the eigenvectors $v^{n\dagger}C(t_0)v^{m} = \delta^{n,m}$ to show that the overlap factors can be obtained from the
eigenvectors using the relation 
\beq
Z^n_i = \sqrt{2m_n}e^{m_nt_0/2}v^{n}_jC_{ji}(t_0).
\eeq{Z_eq}
% These overlap factors are very useful in identify the spin of a given state. 

In \fgn{gerhist}, we show an array of histograms of the normalized overlap factors, $\tilde{Z} = {Z^n_i\over max_n[Z^n_i]}$, of a few operators onto some of the spin identified lower-lying states in each of the lattice irreps. The normalization 
$\tilde{Z} = {Z^n_i\over max_n[Z^n_i]}$ is such that the largest overlap 
factor of that operator across all states is unity.
\bef[tbh]
\centering
\includegraphics[width=18cm,height=11cm]{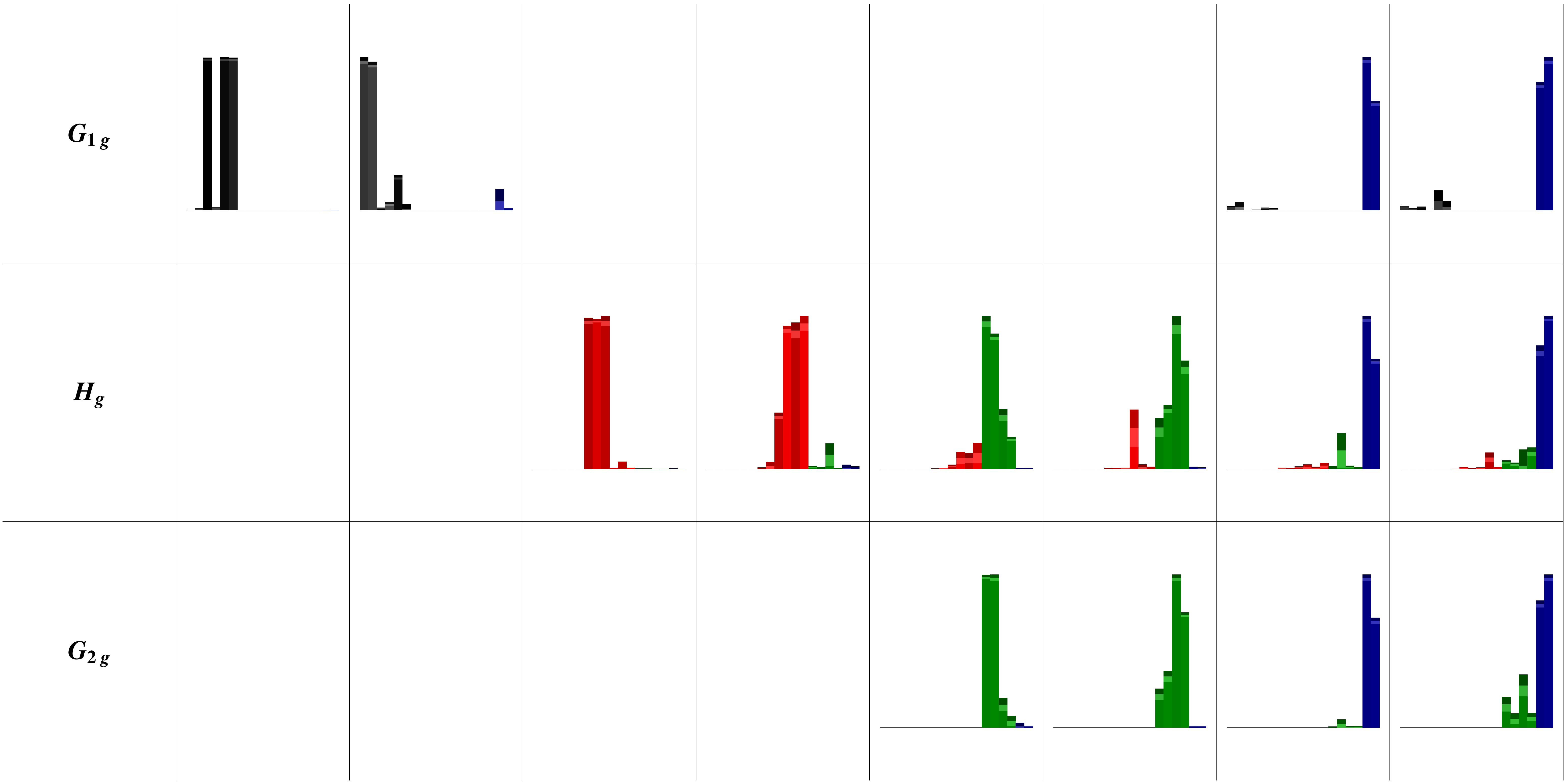}
\caption{Histogram plot showing normalized overlap factor, $\tilde{Z}$, of a few operators onto some of the lower-lying states in each of the lattice irreps.
 $\tilde{Z}$'s  are normalized according to ${Z^n_i\over max_n[Z^n_i]}$, so that the largest value for that operator across all states is equal to unity. 
Top row is for irrep $G_{1g}$, middle one is for $H_g$, and the bottom one is for $G_{2g}$ irrep. Black bars correspond to spin-1/2 operators, red for spin-3/2, green for spin-5/2 and blue represents spin-7/2 operators. 
%Selected operators are also represented by appropriate colour coding. 
Lighter and darker shades on the top of every bar represent the one sigma statistical uncertainty% on negative and positive side of the mean respectively
.}
\eef{gerhist}
%%====================================================

\bef[tbh] \centering
\includegraphics[width=17cm,height=11cm]{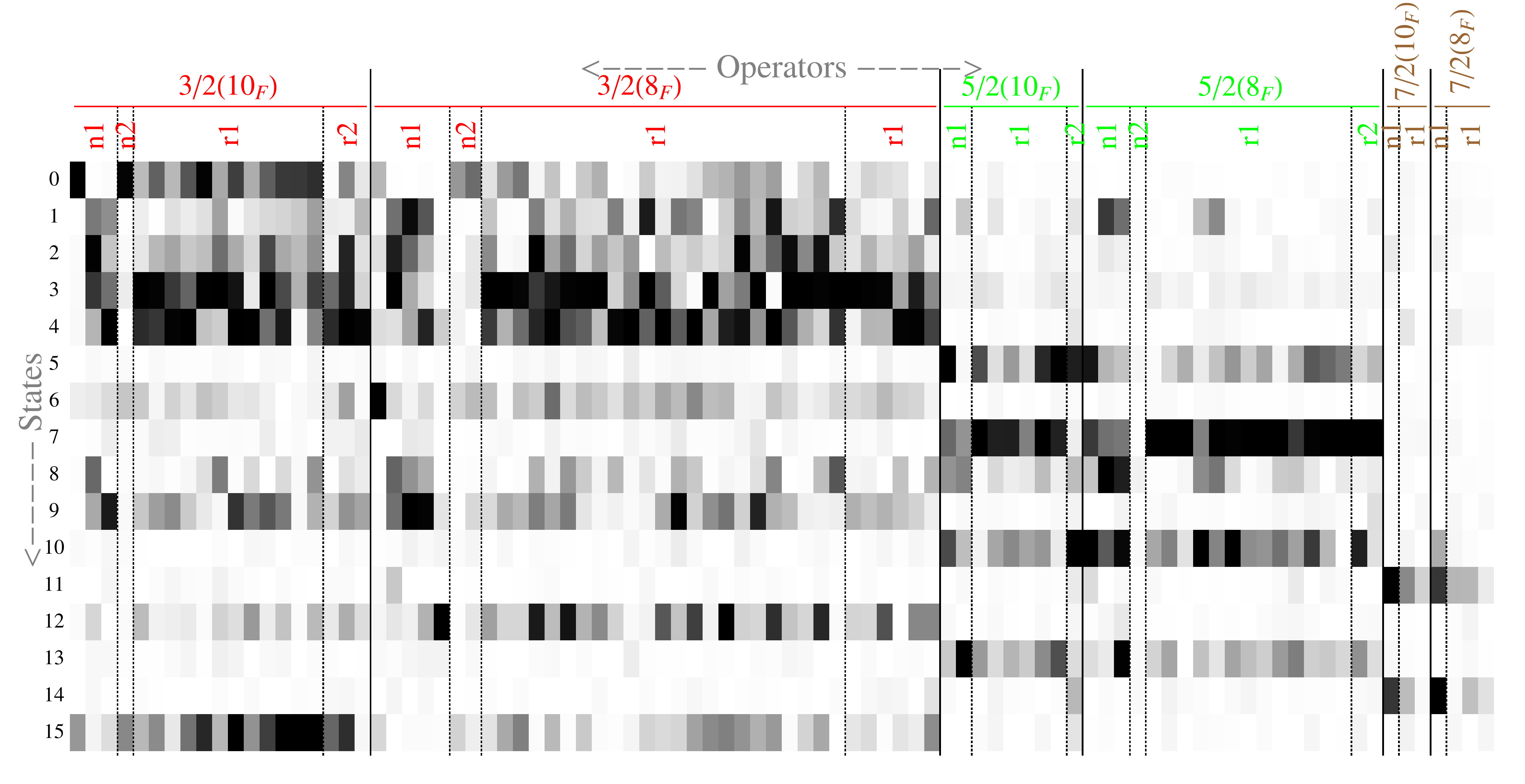}
\caption{``Matrix" plot of the normalized overlap factor,
  $\tilde{Z}^n_i$, of an operator $i$ to a given state $n$, as
defined by \eqn{Z_eq}.  $\tilde{Z}^n_i$ are normalized according to
${Z^n_i\over max_n[Z^n_i]}$, so that for a given operator the largest
overlap across all states is unity.  This plot corresponds to the
extracted states of $\Xi_{cc}$, in the $H_{g}$ irrep. Darker pixels
indicate larger values of the operator overlaps as in
\fgn{rel_ccu_Hg_corr_matplot}. Various type of operators, for example,
non-relativistic (n) and relativistic (r) operators, as well as
non-hybrid (1) and hybrid (2) operators are indicated by column
labels.  In addition, the continuum spins of the operators are shown
by 3/2, 5/2 and 7/2. State 0, the ground state, and excited states 1,
2, 3, 4, 9, 12 and 15 could be identified as $J^P = {3\over 2}^+$
states by the overlap to various types operators according to pixel
strengths. States 5, 7, 8, 10 and 13 could be identified as $J^P =
{5\over 2}^+$ states; similarly, states 11 and 14 could also be
identified as $J^P = {7\over 2}^+$ states.} 
\eef{rel_ccu_Hg_z}
  \bef[tbh] \centering
  \includegraphics[width=17cm,height=11cm]{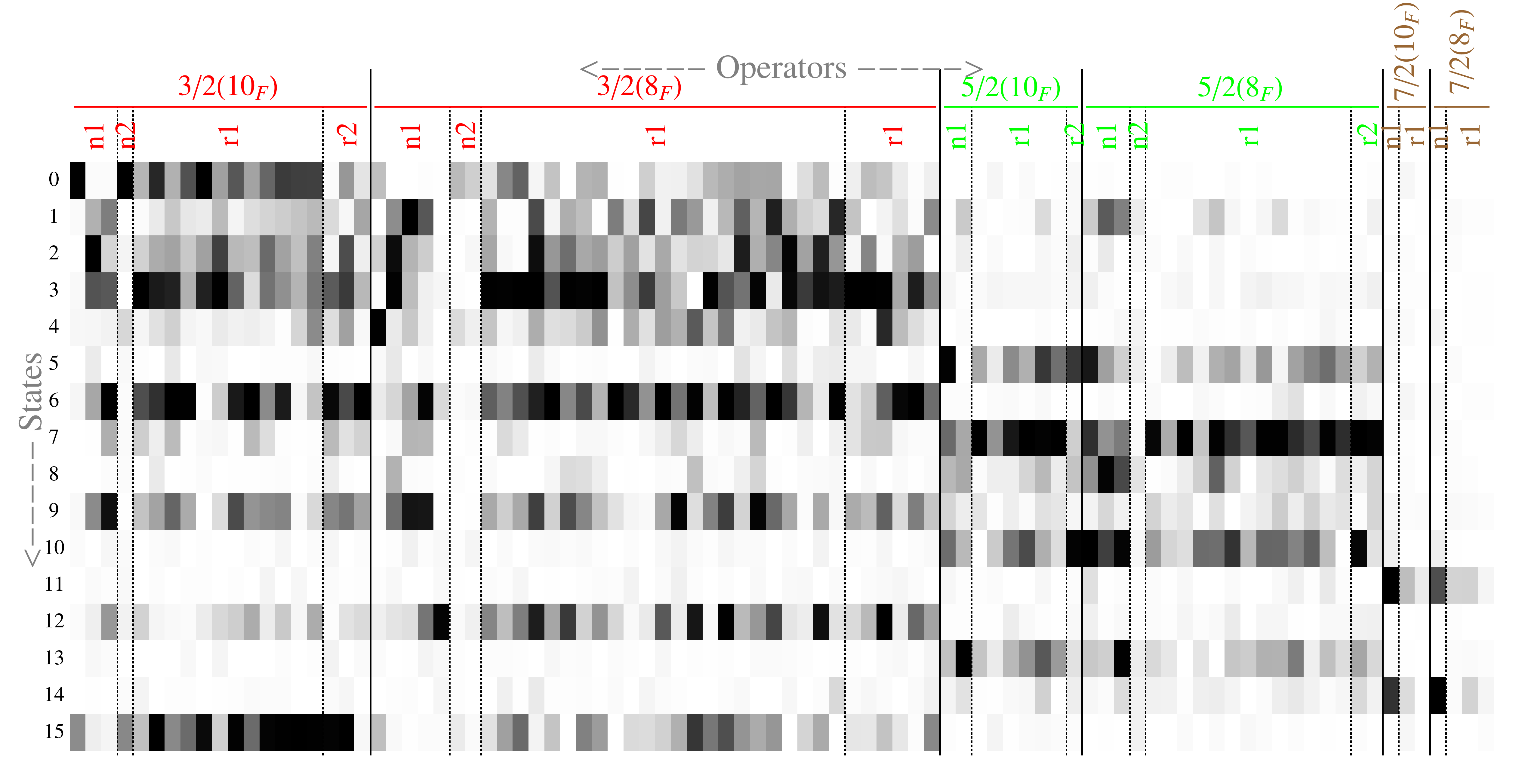}
\caption{Same as \fgn{rel_ccu_Hg_z}, but for $\Omega_{cc}$ baryons in
  the irrep $H_g$.}  \eef{rel_ccs_Hg_z}

As in Ref.\cite{Edwards:2012fx}, a ``matrix'' plot
of the overlap factors can be used to depict the dominant contribution
to the low-lying states from each operator. As examples, \fgn{rel_ccu_Hg_z} and
\fgn{rel_ccs_Hg_z} show such plots of the normalized
overlap factors, $Z^n_i$, of an operator $i$ to a given state $n$, as
defined by \eqn{Z_eq}, for $\Xi_{cc}$ and $\Omega_{cc}$ respectively. All values of $Z^n_i$ are normalized as above. 
%In these plots along the $x$-axis
%we show various operators. Solid lines along columns are used to
%distinguish between operators subduced from different spins while
%dashed lines distinguish operators of different types for a
%given spin. In the $y$-axis states are arranged in the order they
%obtained from the fitting.
%% As mentioned above, these plots represent a pictorial way of 
%% identifying the spin of a fitted state.  
%albeit not the only criteria.  
\fgn{rel_ccu_Hg_z} shows 
that state 0, the ground state, and excited states 1, 2, 3, 4, 9, 12
and 15 are naturally identified as $J^P = {3\over 2}^+$ states while 
states 5, 7, 8, 10 and 13 are identified as $J^P =
{5\over 2}^+$ states and similarly, states 11 and 14 can be
identified as $J^P = {7\over 2}^+$ states. However, in order to
confirm the reliability of the identification of a state with a given
spin greater than 3/2 we compare the magnitudes of overlap
factors of a particular operator from different irreps. This will be
discussed later.
These plots help us to identify the structure of a state
from the types of operators which construct it. For
example, in \fgn{rel_ccu_Hg_z}, the spin-3/2${+}$ ground state has
predominant overlaps from the non-relativistic non-hybrid
as well as relativistic non-hybrid type operators. Similarly, the
states 11 and 14, which are identified as ${7\over 2}^+$ states, have
predominant overlaps to non-relativistic non-hybrid operators. Strong
hybrid content was observed for a number of states by identifying
their strong overlaps with hybrid interpolating operators.

To identify the spin parity of a state we followed the same method
detailed in \cite{Dudek:2007wv} and used for light mesons \cite{Dudek:2009qf, Dudek:2010wm, Dudek:2010ew}, baryons \cite{Edwards:2011jj, Edwards:2012fx}, charm mesons \cite{Liu:2012ze} as well as heavy-light mesons \cite{Moir:2013ub} calculations. To identify a spin that subduces into a single irrep is relatively 
straightforward and studying the overlap factors used in histogram and 
matrix plots we can identify spin-$\frac12$ and spin-$\frac32$
states. For spin-$\frac52$ and spin-$\frac72$ states, which
are subduced into multiple irreps overlap factors from different irreps must be 
compared.  As the continuum limit is approached, the factors of a given 
continuum operator to a particular state obtained from various subduced irreps 
should be the same. For a fine lattice spacing 
these overlap factors should thus be close to each other. For example, the 
spin-7/2 continuum operator, $(3/2^{+})_{1,S} \otimes D^{[2]}_{L=2S}$, can be
subduced to irrep $G_{1g}$, $H_{g}$ as well as to $G_{2g}$. 
%However,
%it is expected that, at a finite lattice spacing, for a particular
%state which is near degenerate over these irreps, overlap factors for
%this operator to that state would also be degenerate. 
%However, it is expected that, 
%at lattice spacings that are ``close'' to the continuum, the overlap factors for these
%lattice operators to the same $\frac72^+$ state would be degenerate.
This near degeneracy of overlap factors can be used to identify this state as
spin-$\frac72^{+}$ state. In \fgn{Z_compare} we compare a selection of
$Z$-values for states conjectured to be $J = \frac52^+$ (top two
plots), $\frac72^+$ (middle two plots), $\frac52^-$ (bottom left) and
$\frac72^-$ (bottom right) which appear over multiple lattice irreps.
The continuum operators considered are noted along the lower edge of each
plot. $Z$-values obtained for a given operator but from different
irreps are found to be consistent with each other which helps us to
identify the spin of these given states.

After identifying the spin of a state with matching overlap
factors we check whether the energy of this state determined over different 
irreps also matches. To achieve this check, the corresponding principal 
correlators across the irreps are simultaneously fit to a single energy. 
In \fgn{joint_correlator}, these joint fits are shown for the principal
correlators obtained from three different irreps for a spin-7/2$^{+}$ state
(top plot) and from two different irreps for a spin-5/2$^{+}$ state (bottom
plot).

%\FloatBarrier

%\clearpage
%\begin{protrait}

\bef[tbp]
\centering
\vspace*{-0.1in}
\includegraphics[width=8cm,height=6.5cm]{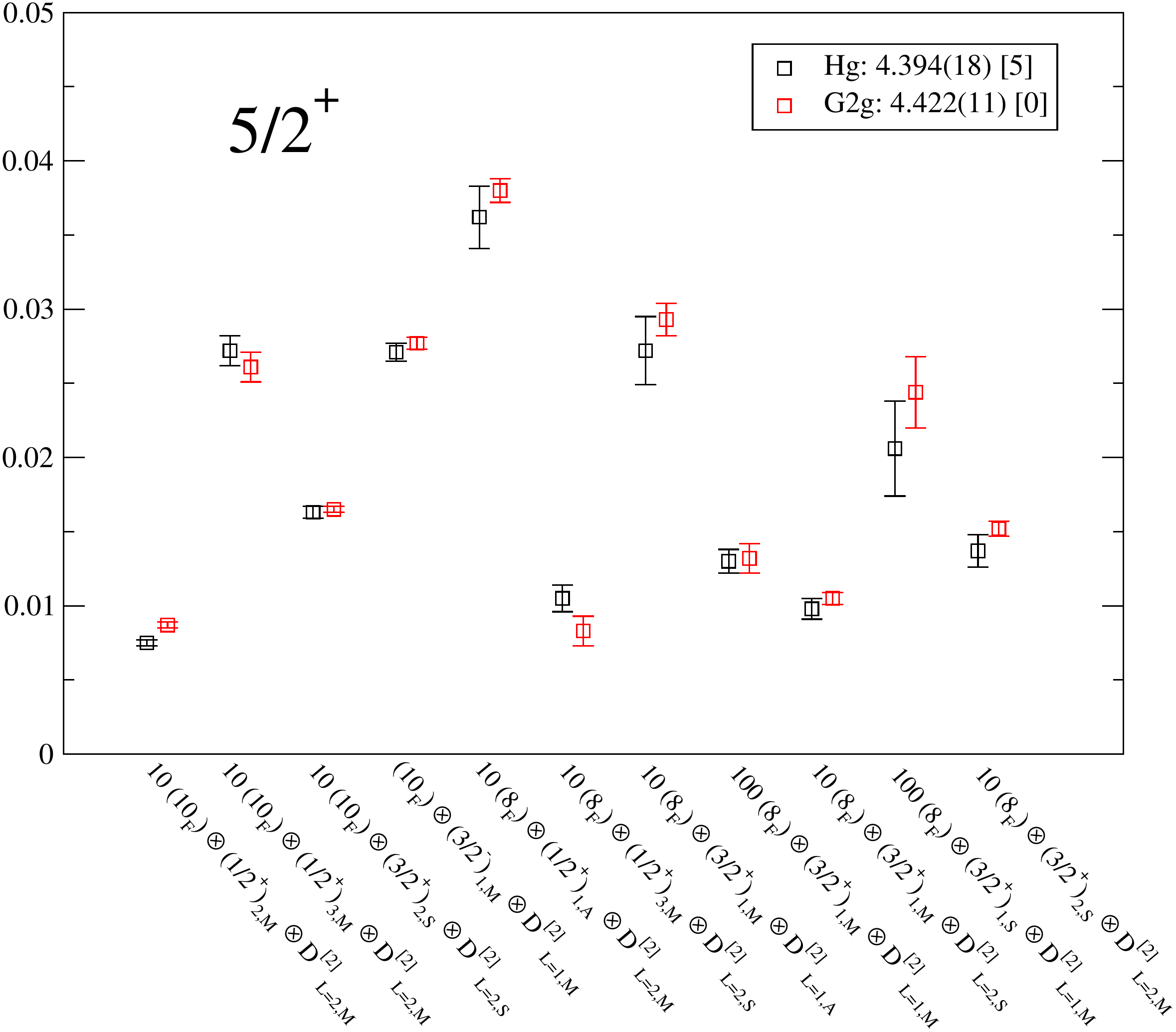}
\hspace*{0.2in}
\includegraphics[width=8cm,height=6.5cm]{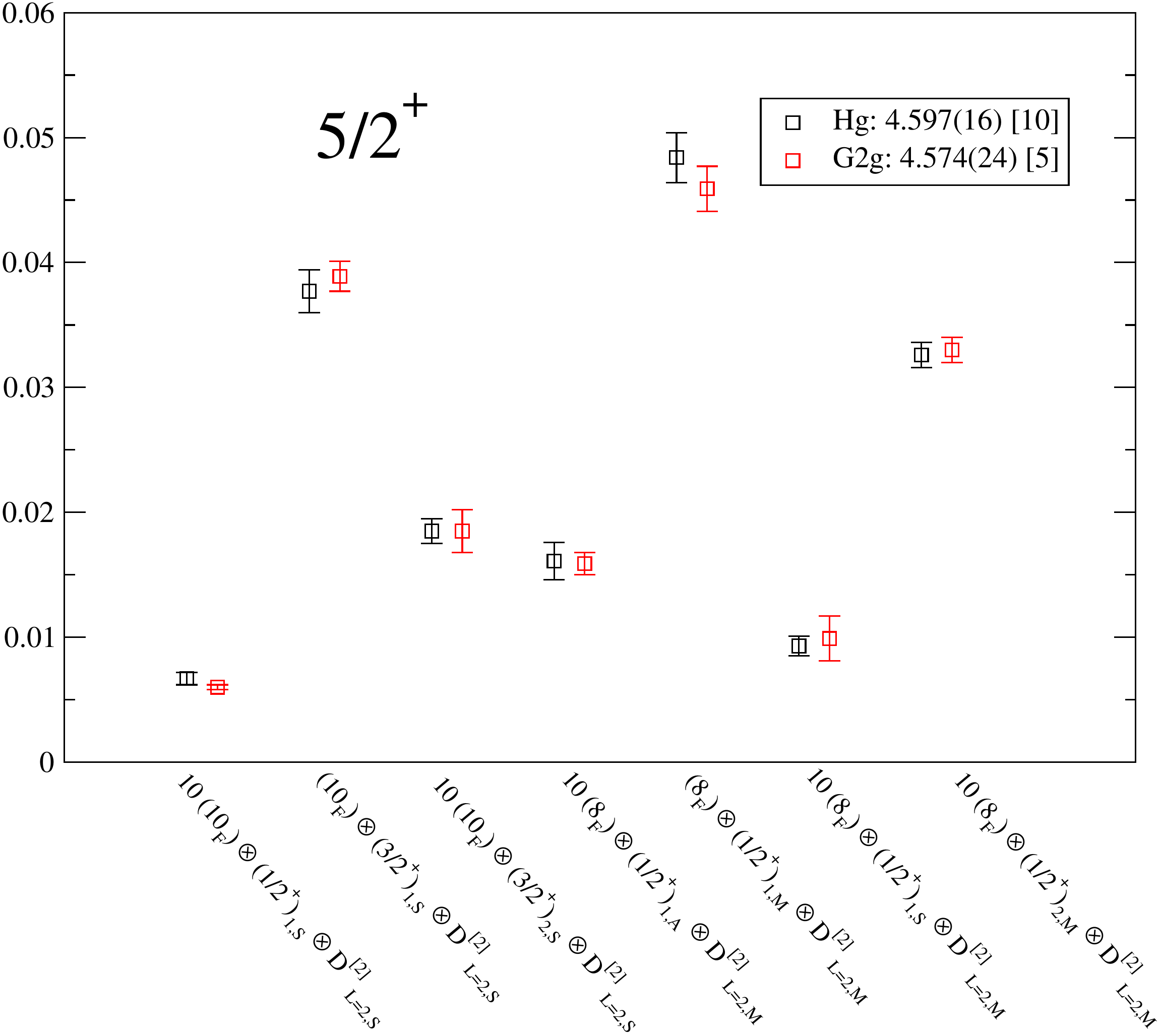}\\
\vspace*{0.1in}
\includegraphics[width=8cm,height=6cm]{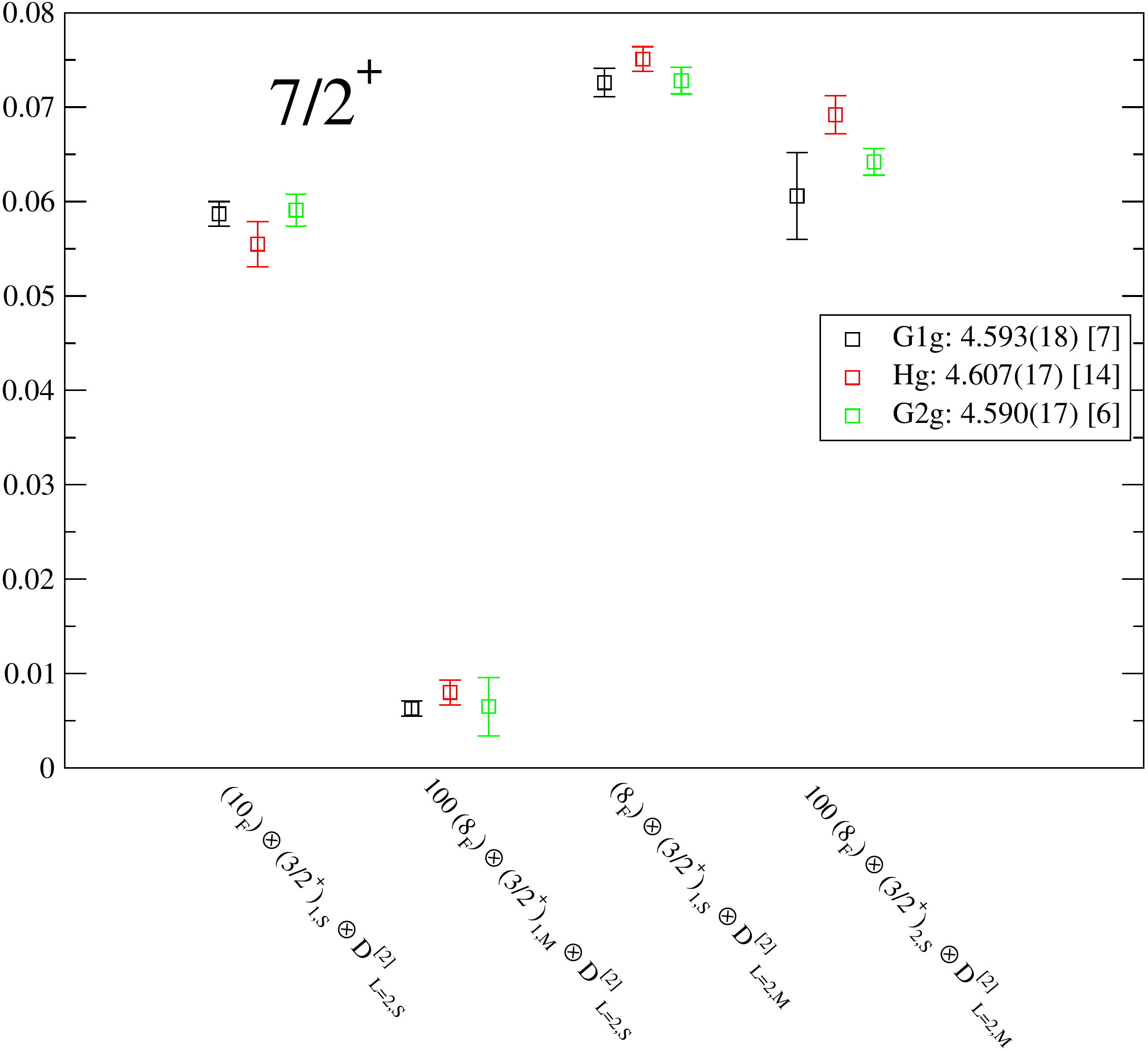}
\hspace*{0.2in}
\includegraphics[width=8cm,height=6cm]{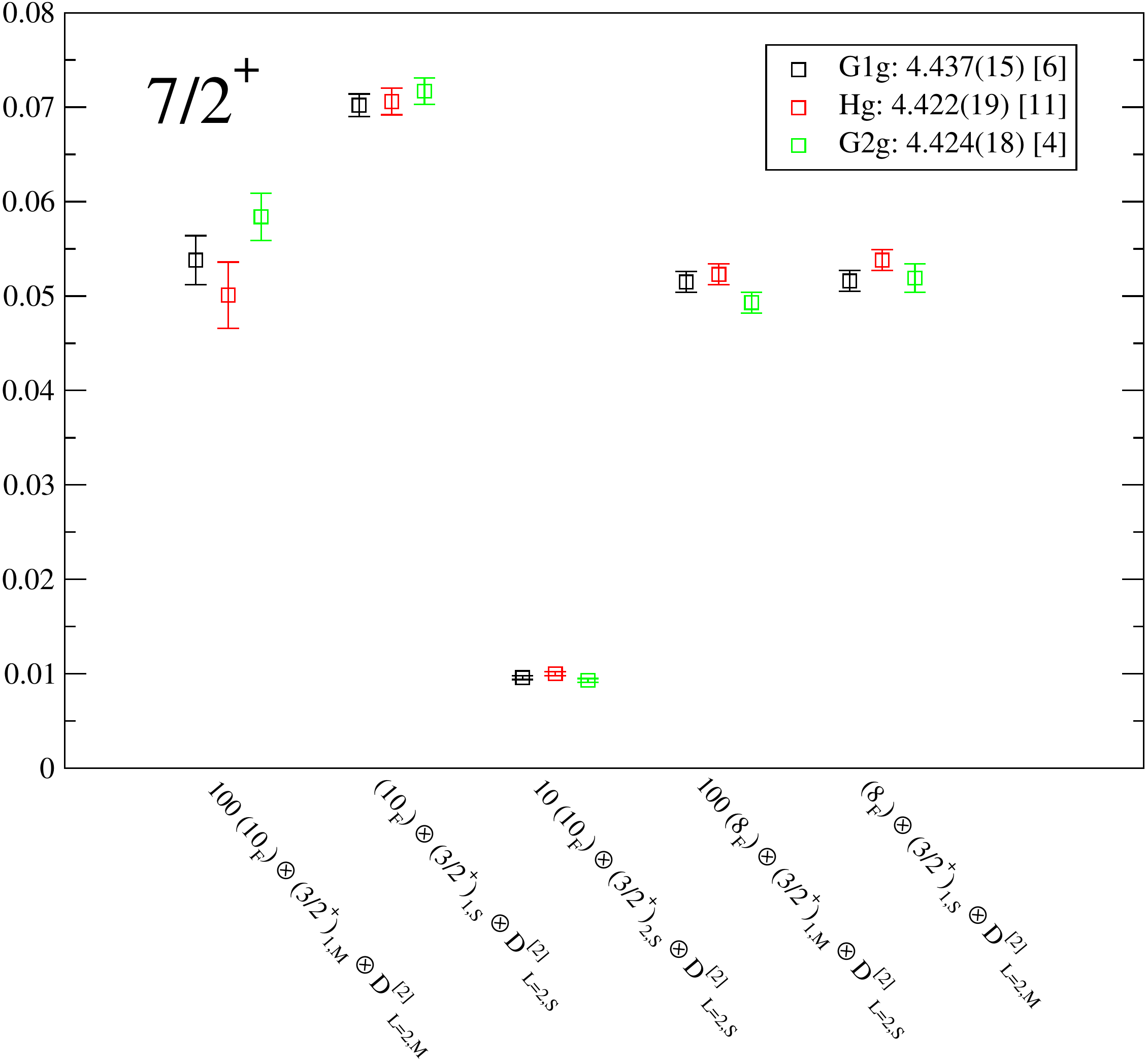}\\
\vspace*{0.1in}
\includegraphics[width=8cm,height=6cm]{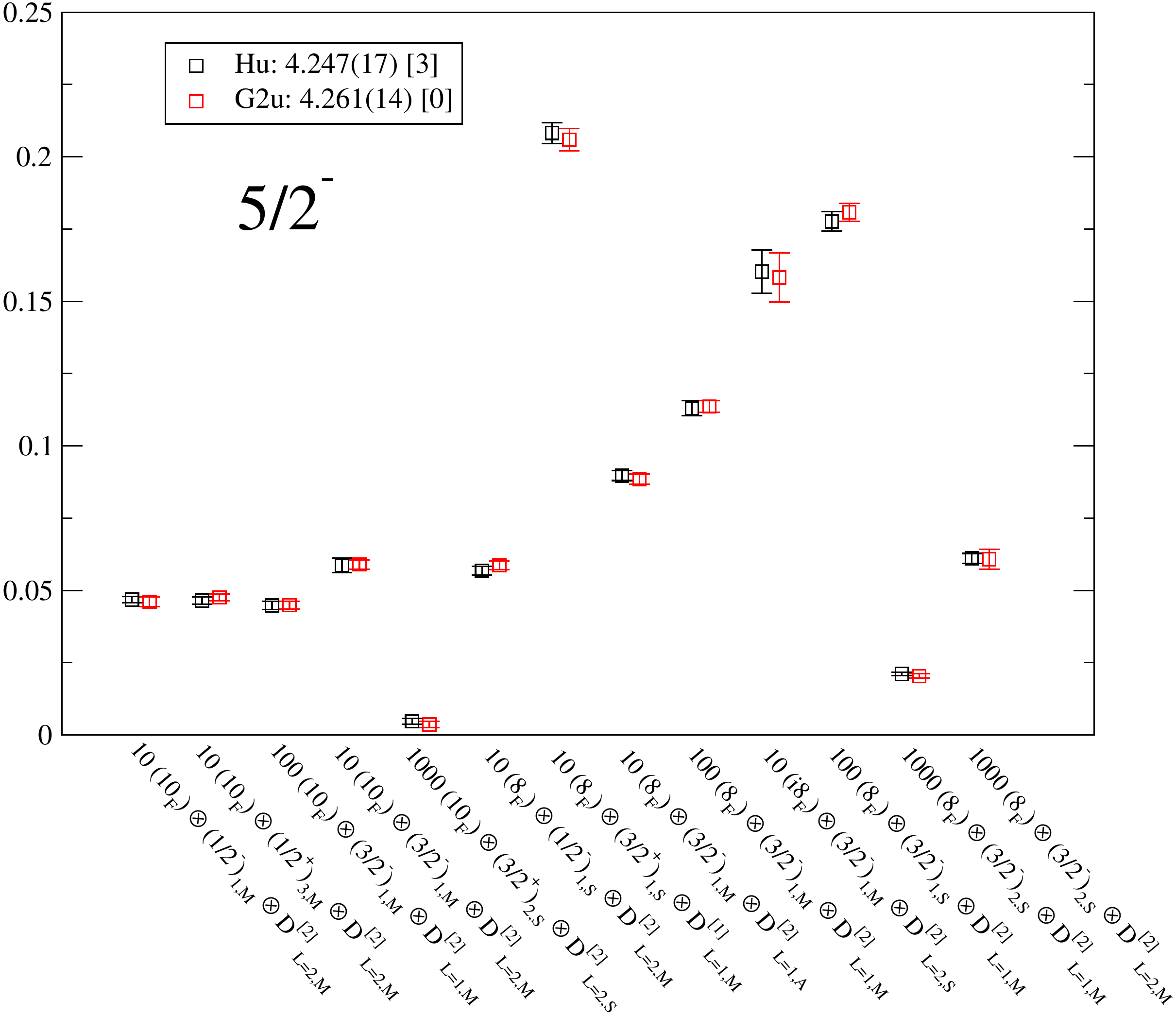}
\hspace*{0.2in}
\includegraphics[width=8cm,height=6cm]{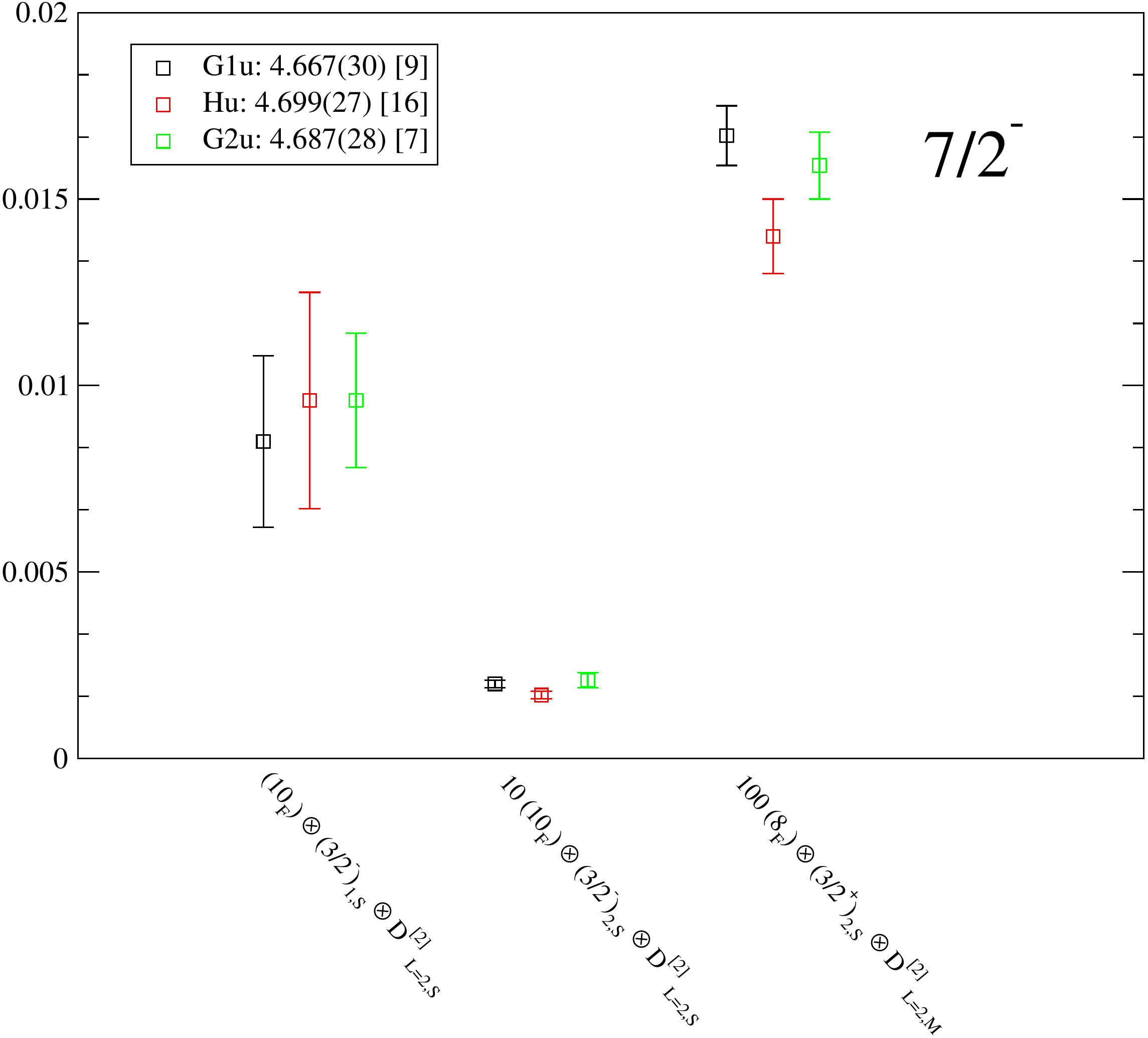}
\caption{A selection of $Z$-values for states conjectured to be $J = \frac52^+$ (top two plots), $\frac72^+$ (middle two plots), $\frac52^-$ (bottom left) and $\frac72^-$ (bottom right). The operators used are mentioned at the bottom of each plot. $Z$-values obtained for a given operator, but from different irreps, are found to be consistent.}
\vspace*{1in}
\eef{Z_compare}

\bef[tbh]
\centering
\includegraphics[scale=0.9]{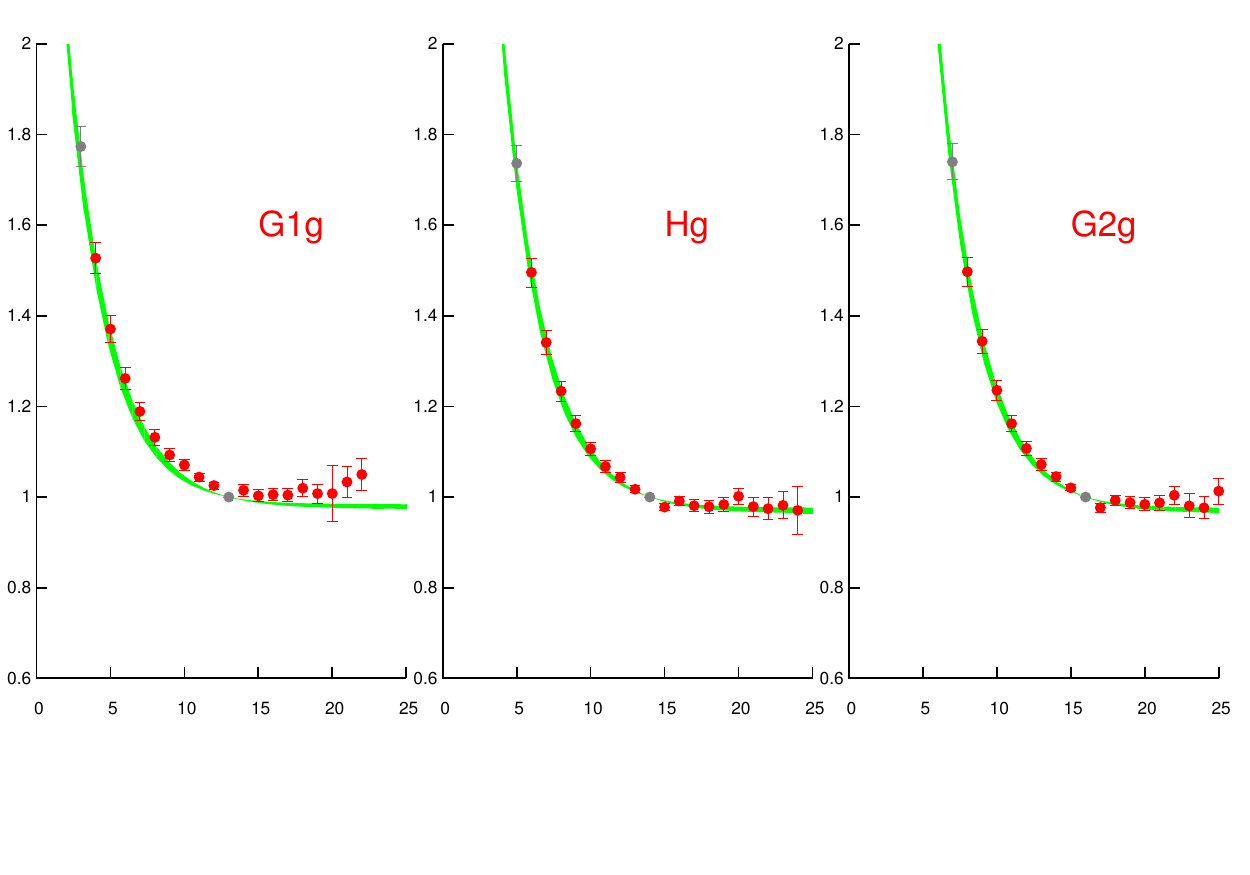}\\
\vspace*{-0.7in}
\includegraphics[scale=0.9]{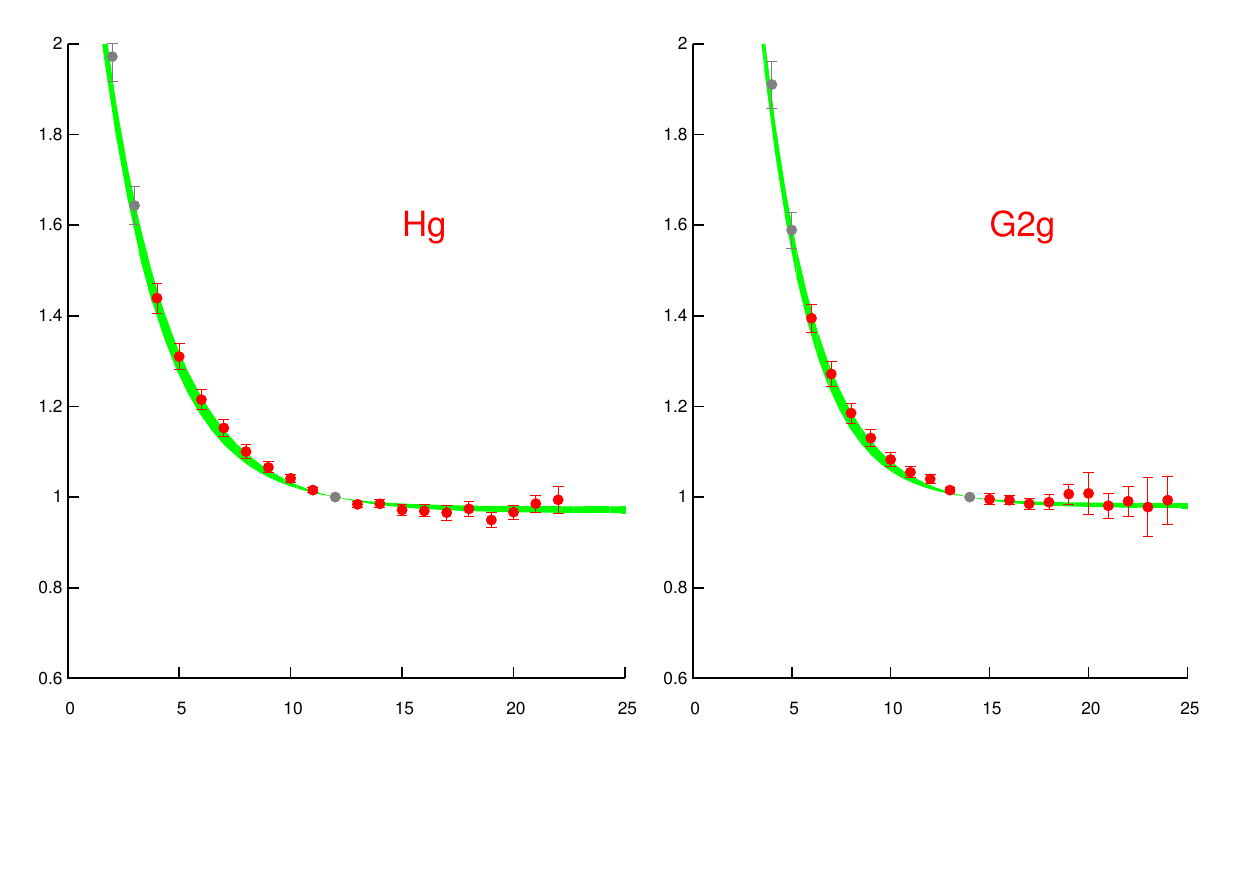}
\vspace*{-0.4in}
\caption{Joint fit of the three principal correlators : top figures -- for a representative spin-$\frac72^{+}$ state from three different irreps, and bottom figures -- for a spin-$\frac52^{+}$ state from two different irreps. Joint fits with $\chi^2/dof \sim 1.9$ (for both cases) provide the masses of these states.}
\eef{joint_correlator}

%==================================================================
%==================================================================
%==================================================================
%==================================================================
\section{Results}
In this section, our results for the doubly-charmed baryons spectra with
spins up to $\frac72$ and with both parities are presented in terms of energy 
splittings. In general, this tends to reduce systematic uncertainties in 
lattice calculations, including those from the scale setting procedure used.
For this work, the charm quark mass parameter in the lattice action was 
determined by ensuring the physical value and lattice estimate of the 
$\eta_c$ meson mass~\cite{Liu:2012ze} agreed, once the lattice spacing is 
determined using the $\Omega$-baryon. The $\eta_c$ meson has the same 
number of charm
quarks as the doubly-charmed baryons. Hence, we show the spectra of
$\Xi_{cc}$ and $\Omega_{cc}$ baryons with the mass of the $\eta_c$
meson subtracted. These energy splittings expose the binding energies of the 
extra light and strange %$u$ and $s$
 quarks inside these baryons. A few energy splittings such as the hyperfine 
splittings between the extracted states are also computed.

As mentioned earlier, the experimental status of doubly heavy
baryon discovery is uncertain. It is thus important to compare the ground
state spectra of the two doubly heavy baryons obtained from different
calculations. In \fgn{groundstate_ccu}, we show the ground state
results of $J^{P} = \frac12^+, \frac32^+,\frac12^-$ and $\frac32^-$ of
$, \Xi_{cc}$ baryons, obtained in this work, along with the only
experimental (SELEX) result and other lattice as well as various model
results. In \fgn{groundstate_ccs}, we show similar results for
$\Omega_{cc}$ baryons. Our results are at pion mass 391 MeV, results
for ETMC \cite{Alexandrou:2012xk}, PACS-CS~\cite{Namekawa:2013vu},
Bali {\it et. al}~\cite{Bali:2012ua}, and Briceno {\it et. al}~\cite{Briceno:2012wt} are extrapolated to the physical pion
mass, while ILGTI \cite{Basak:2012py} results are at pion mass 390
MeV.  While the lattice spacing in the temporal direction ($a_t$) for
this work is 0.0351 fm, for ETMC, $a_t = 0.056$ fm, for PACS-CS, $a_t =
0.0899$ fm, for Bali {\it et. al}, $a_t = 0.0795$ fm and for ILGTI, $a_t
= 0.0582$ fm. Results for Briceno {\it et. al} are extrapolated to 
the continuum limit.

In \fgn{split_ccu} and \fgn{split_ccs} we show the spin identified full spectra, up to $J={7\over2}$ extracted from our lattices of $\Xi_{cc}$ and $\Omega_{cc}$ baryons, respectively. The states inside
the magenta boxes are those with relatively large overlap onto
non-relativistic operators and the states with thick borders
corresponds to the states with strong hybrid content as defined in Ref.~\cite{Dudek:2012ag}.

Note that in the lowest two positive parity bands and the
lowest negative parity band, the number of states for each spin agrees
with the expectation shown in \tbn{n_operators}. That table gives 
the number of allowed quantum numbers by $SU(6)\times O(3)$
symmetry for operators with up to two derivatives (D). For $J^{P} =
\frac12^{+}$ and $J^{P} = \frac32^{+}$, the number of allowed quantum
numbers is 7 (1 from D = 0 and 6 from D = 2) and 9 (1 from D = 0 and 8
from D = 2) respectively. For other positive parity quantum numbers
with $J^{P} = \frac52^{+}$ and $\frac72^{+}$ these numbers are 5 and 2
respectively. The left sides of \fgn{split_ccu} and
\fgn{split_ccs} show positive parity states. The numbers of
states in the lowest two bands (first two and inside box) match
exactly with the allowed quantum numbers mentioned above. Note that while we 
use the full set including both non-relativistic and relativistic operators 
not in \tbn{n_operators}, we still obtain the same number of states
allowed with only non-relativistic operators.  Similarly, for negative
parity with $J^{P} = \frac12^{-}\,, \frac32^{-}\,, \frac52^{-}\,,
\frac72^{-}$, the allowed number of quantum numbers are 3, 3, 1 and 0,
respectively. On the right sides of \fgn{split_ccu} and
\fgn{split_ccs}, the lowest band (inside box) also has exactly the
same number of states.  This agreement of the number of low lying
states between the lattice spectra obtained in this work and the
expectations based on non-relativistic quark spins implies a clear
signature of $SU(6)\times O(3)$ symmetry in the spectra. Such
$SU(6)\times O(3)$ symmetric nature of spectra was also observed in
Ref.~\cite{Edwards:2012fx,Padmanath:2013zfa}. Note there
are no negative parity spin-7/2 state in that table and the
negative parity spin-7/2 state is obtained from the inclusion of relativistic operators. 
We are also able to identify one
state with strong overlap onto a hybrid operator. Though there are
more quantum numbers accessible with the operators in \tbn{n_operators} with 
hybrid structures, we could not clearly identify those. This is because, as
noted in Ref.~\cite{Edwards:2012fx}, it is not meaningful to interpret the 
higher excited states in terms of $SU(6)\times O(3)$ symmetry.

\bef[!t]
\centering
%\vspace*{-0.1in}
\includegraphics[scale=0.55]{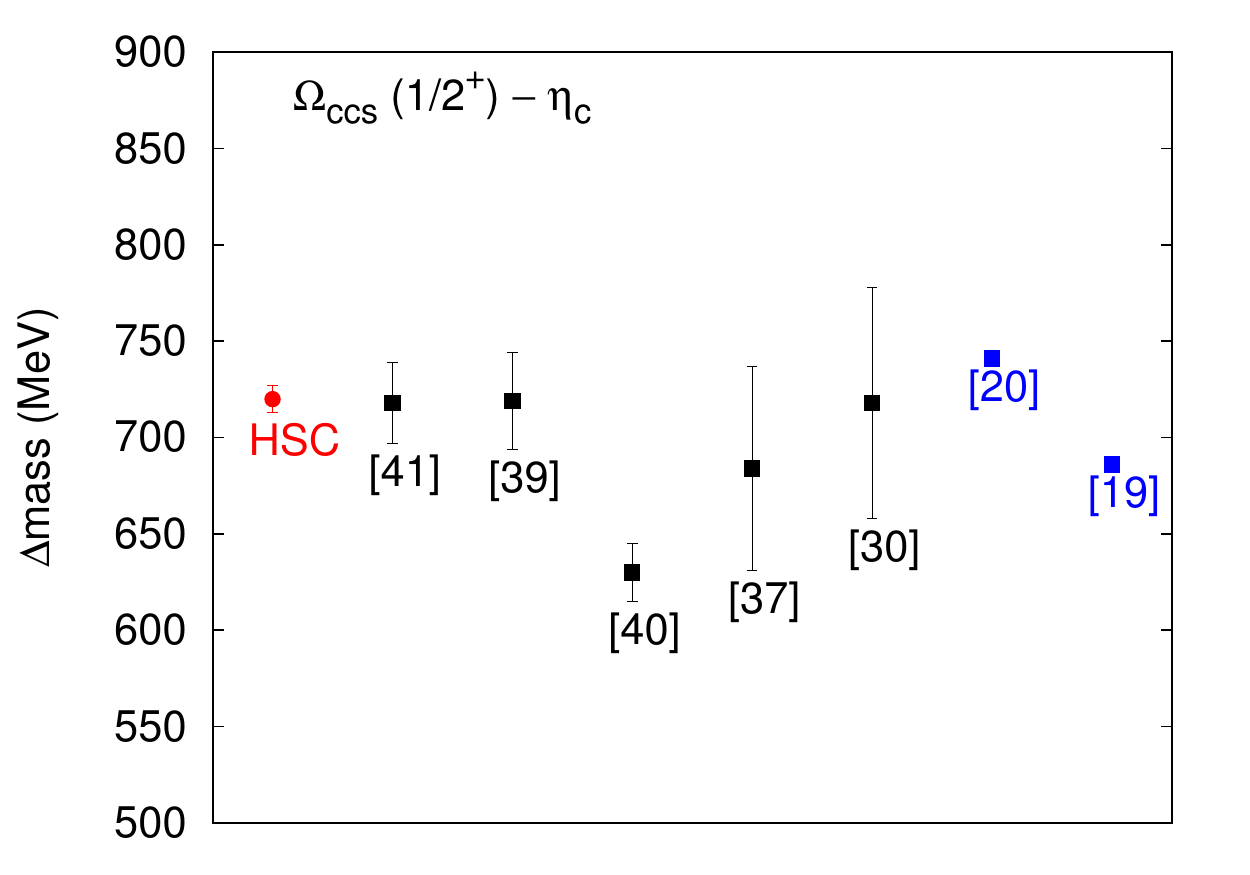}
%\hspace*{0.1in}
\includegraphics[scale=0.55]{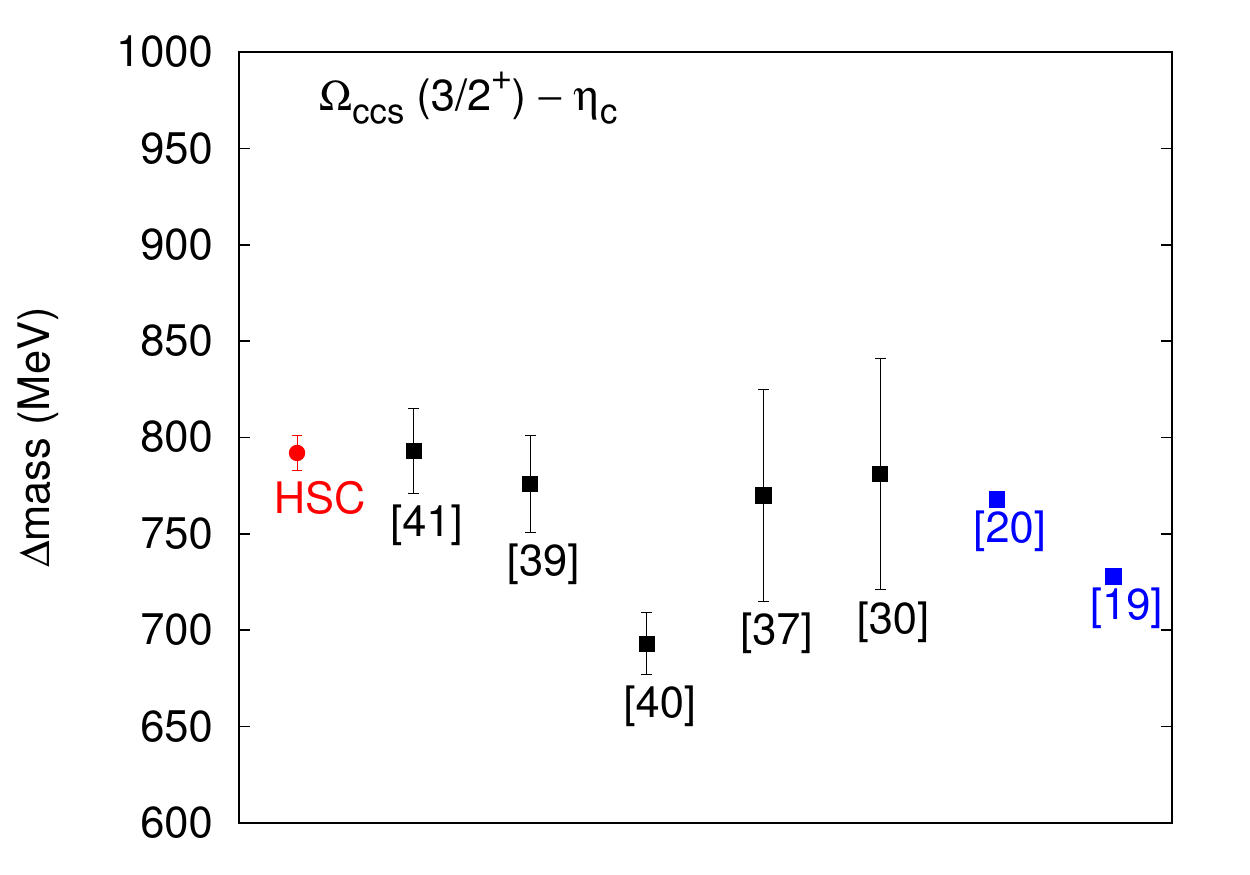}\\
%\vspace*{0.1in}
\includegraphics[scale=0.55]{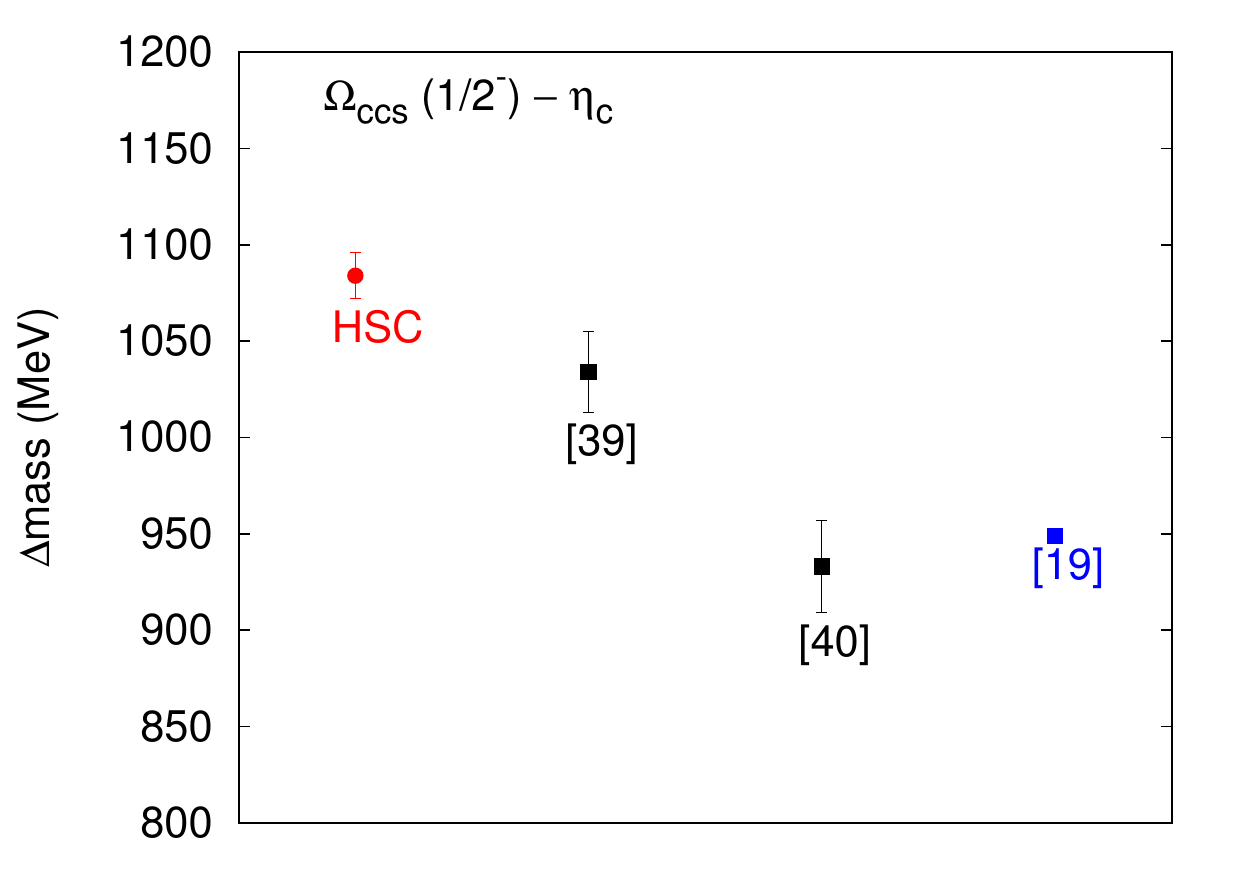}
%\hspace*{0.1in}
\includegraphics[scale=0.55]{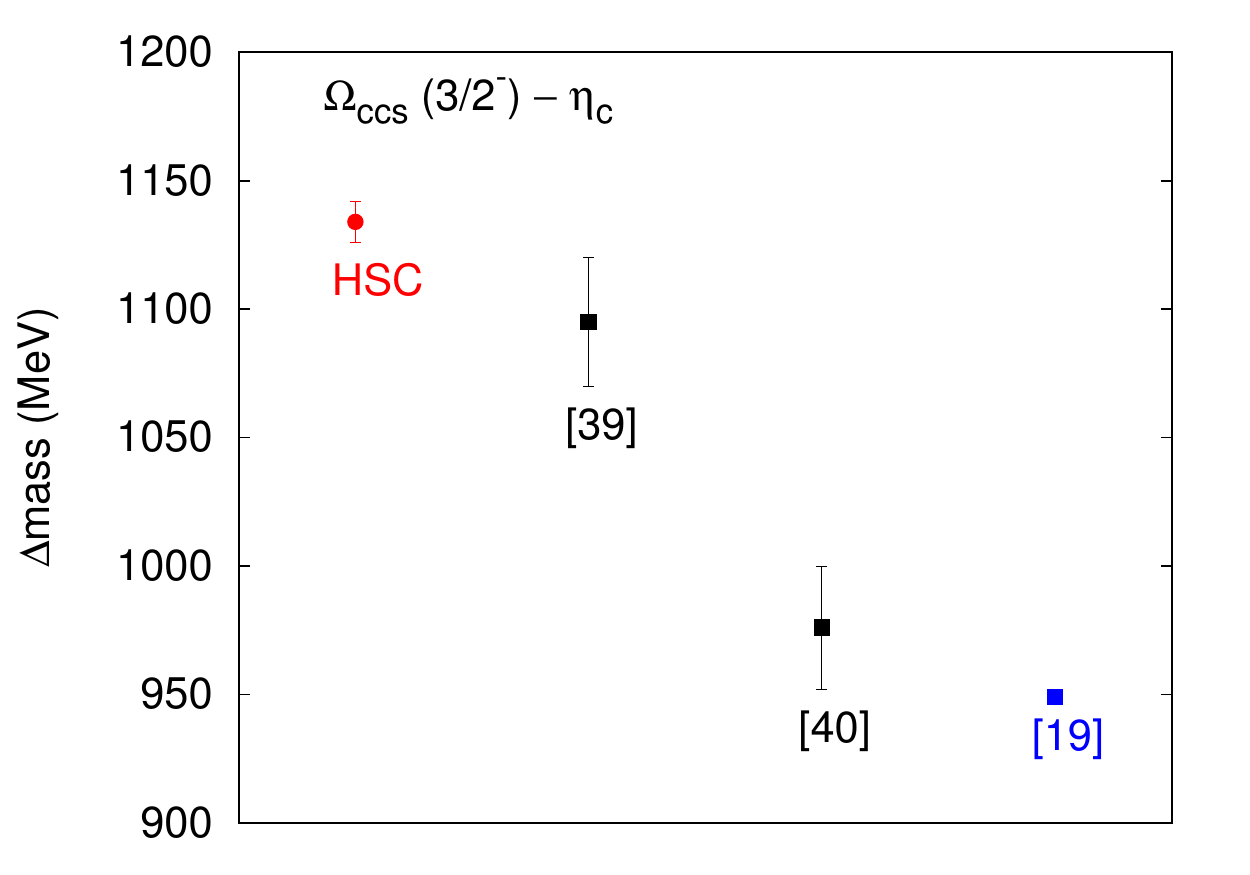}
%\includegraphics[scale=]{0.5}{ccs_1o2p.eps}
%%\hspace*{0.1in}
%\includegraphics[scale=]{0.5}{ccs_3o2p.eps}\\
%%\vspace*{0.1in}
%\includegraphics[scale=]{0.5}{ccs_1o2n.eps}
%%\hspace*{0.1in}
%\includegraphics[scale=]{0.5}{ccs_3o2n.eps}\\
\caption{Ground state masses of spin-1/2 and spin-3/2 doubly-charmed
  $\Omega$ baryons as a splitting from $\eta_c$ meson mass. 
  Our results are shown by the red filled circle (HSC). 
  Other lattice as well as model results are also shown.}
\hspace*{0.1in}
\eef{groundstate_ccs}
%\fgn{groundstate_ccu}
%\clearpage

\bef[b!]
\centering
%\hfill 
%\vspace*{-0.1in}
%includegraphics[width=8cm,height=6cm,angle=90]{fig/ccs_1o2p.eps}
\includegraphics[scale=0.55]{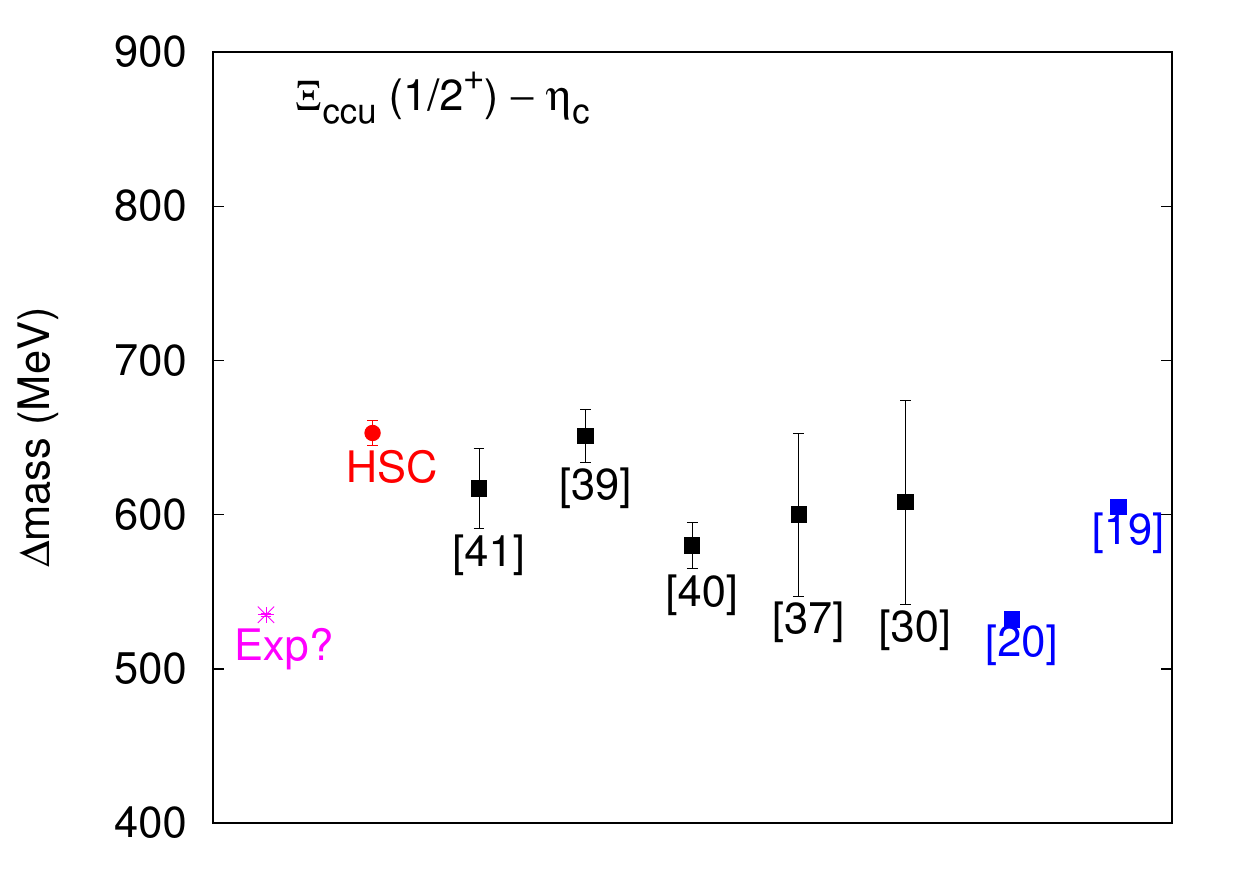}
%\includegraphics[scale=]{0.5}{ccs_1o2p.eps}
%\hspace*{0.1in}
\includegraphics[scale=0.55]{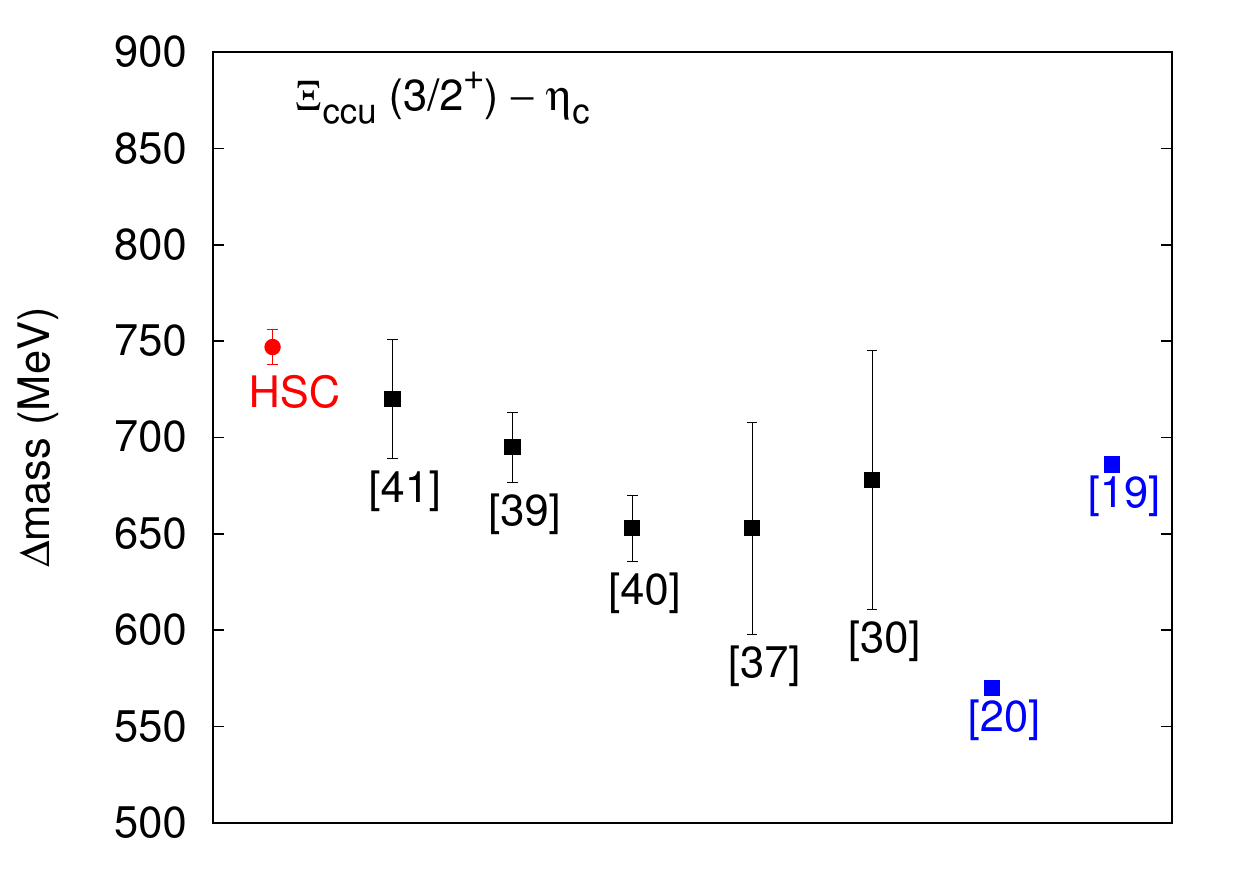}\\
%\vspace*{0.1in}
\includegraphics[scale=0.55]{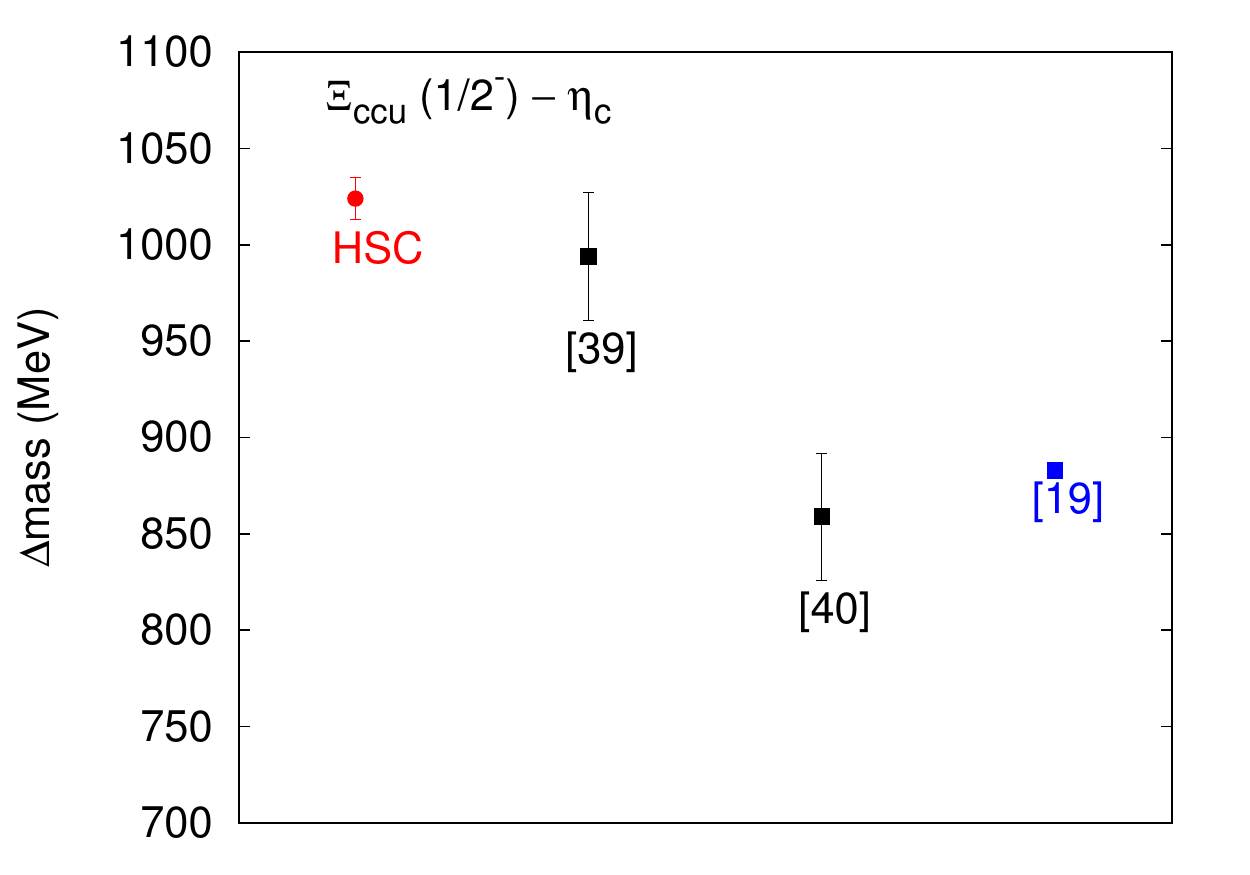}
%\hspace*{0.1in}
\includegraphics[scale=0.55]{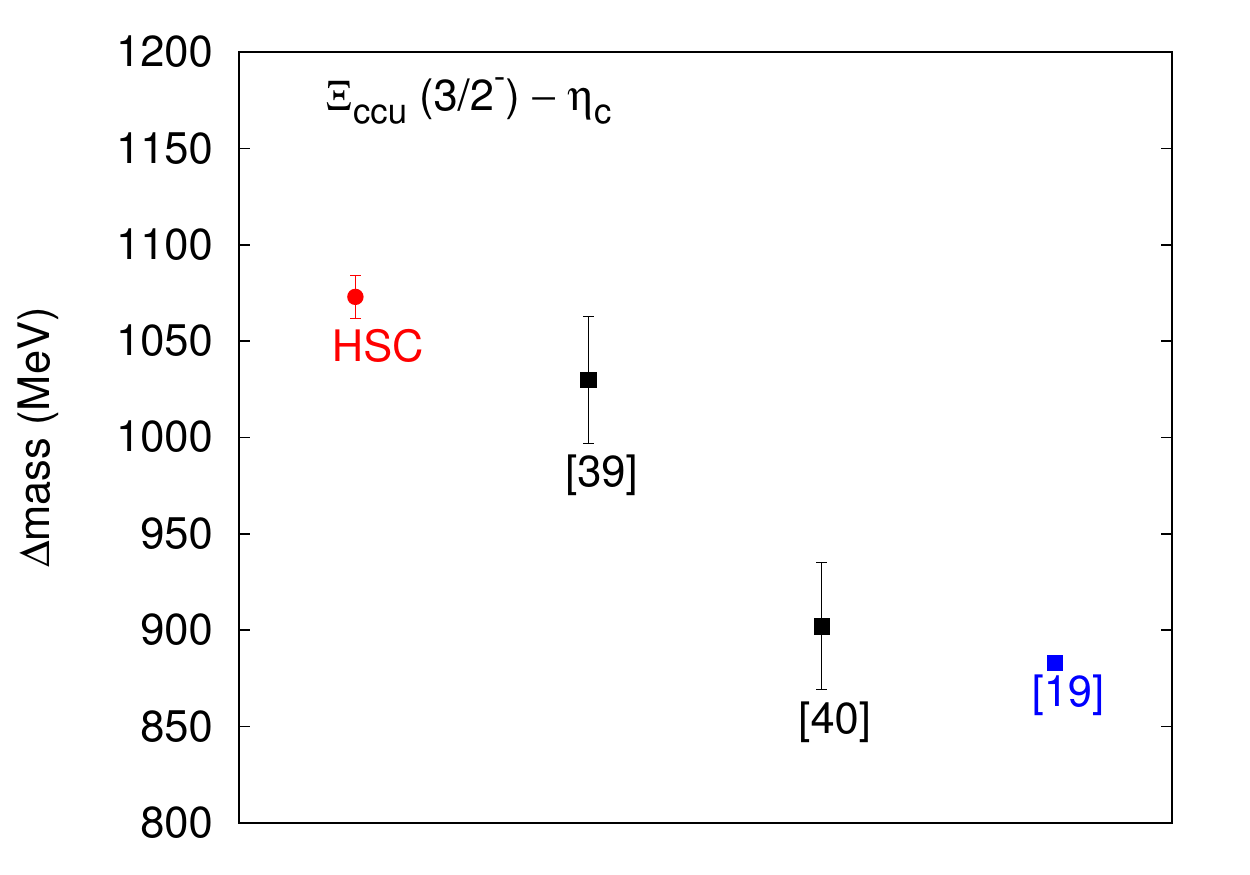}

\caption{Ground state masses of spin-1/2 and  spin-3/2 doubly-charmed $\Xi_{cc}$ baryons as a splitting from $\eta_c$ meson mass. 
  Our results are shown by the red filled circle (HSC). 
  Our $u$ quark mass is not physical and at pion mass 392 MeV,
  results for ETMC \cite{Alexandrou:2012xk}, PACS-CS
  \cite{Namekawa:2013vu}, Bali {\it et. al}~\cite{Bali:2012ua}, and
  Briceno {\it et. al} \cite{Briceno:2012wt} are extrapolated to the
  physical pion mass, while ILGTI \cite{Basak:2012py} results are at
  pion mass 390 MeV and 340 MeV respectively.}
\eef{groundstate_ccu}

\bef[t!]
\centering
\includegraphics[scale=0.48]{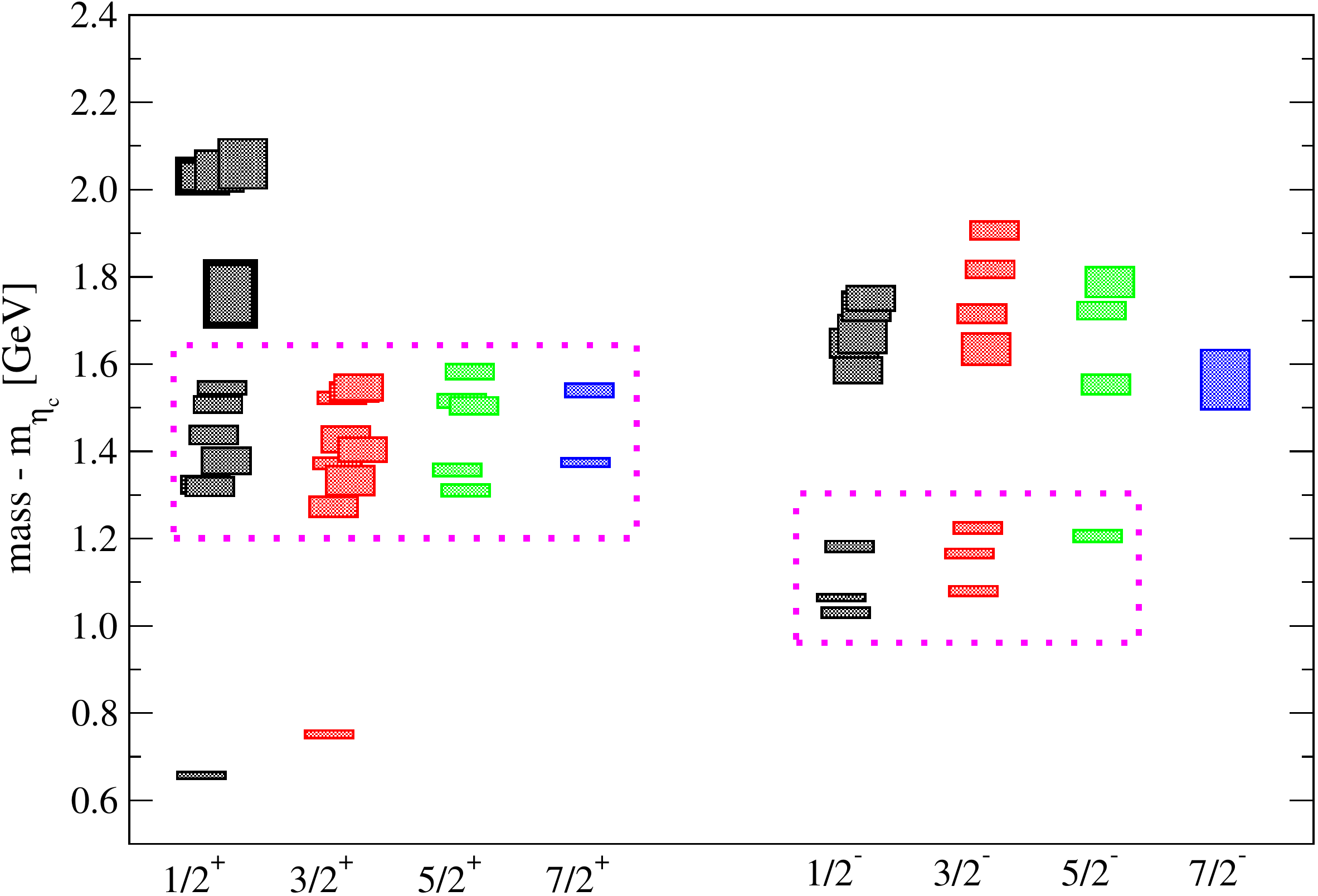}
\caption{Spin identified $\Xi_{cc}$ baryon spectra for both parities and with spin up to 7/2. Energy splittings of the
  $\Xi_{cc}$ baryons from the mass of $m_{\eta_c}$ meson, which has same number of
  charm quarks, are shown here. The states inside the pink boxes are
  those with relatively larger overlap to non-relativistic operators
  and the states with thick borders corresponds to the states with
  strong hybrid content. The number of states inside these boxes
  matches with the expectations based on non-relativistic quark spins,
  as shown in \tbn{n_operators}. This agreement of the number of low
  lying states between the lattice spectra obtained in this work and
  the expectations based on non-relativistic quark spins implies a
  clear signature of $SU(6)\times O(3)$ symmetry in the spectra.}
\vspace*{0.2in}
\eef{split_ccu}
\bef[b!]
\centering
%\vspace*{-0.4in}
\includegraphics[scale=0.48]{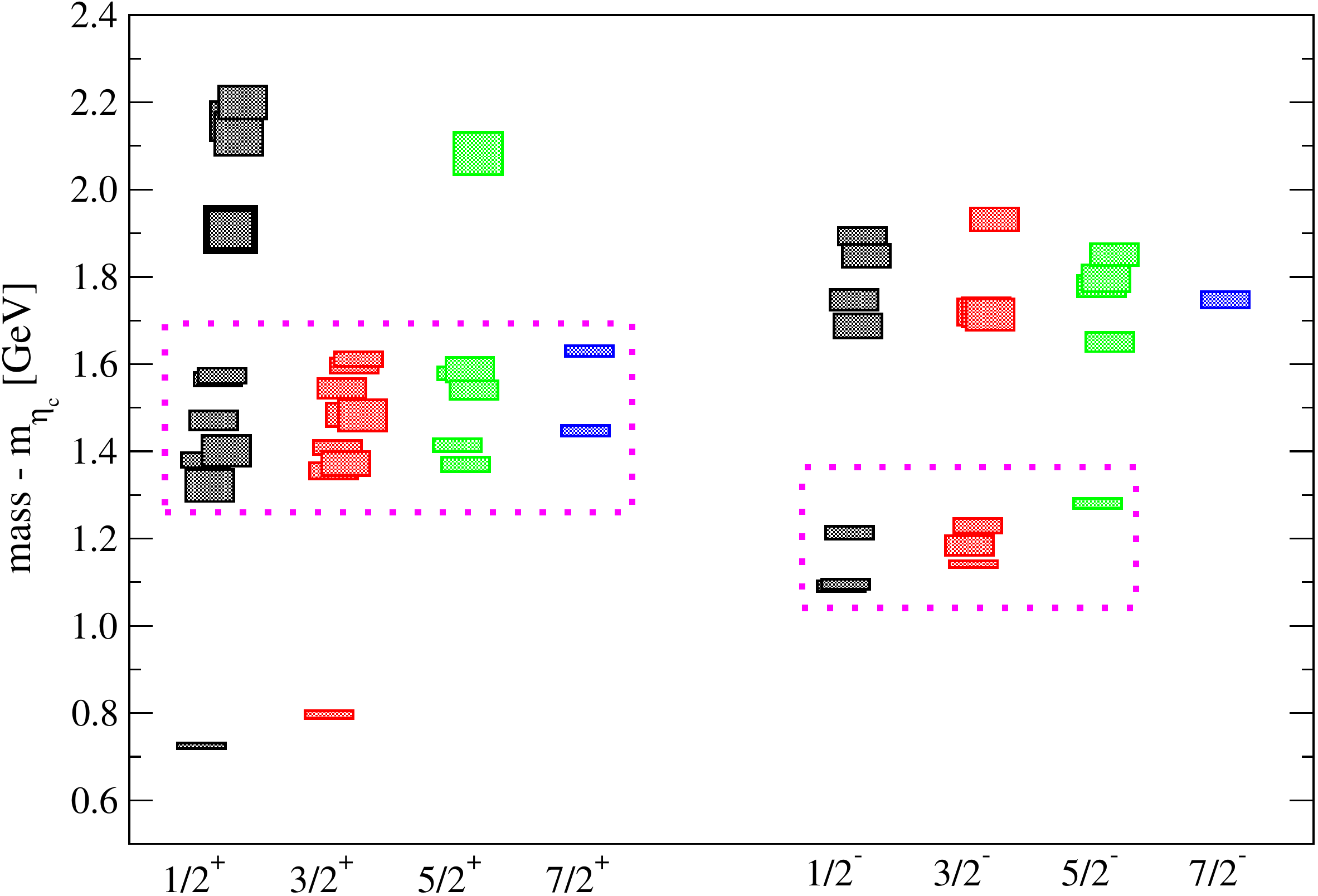}
\caption{Spin identified spectra of $\Omega_{cc}$ baryons for both parities and with spin up to 7/2. Energy splittings of
  the $\Omega_{cc}$ states from the mass of ${\eta_c}$ meson, which has same
  number of charm quark, are shown here. All other details are same as in 
\fgn{groundstate_ccu}.}
%\vspace*{0.32in}
\eef{split_ccs}

\fgn{split_p_ccu_ccs_d_ds} shows the extracted spectra with
a different subtraction. The doubly-charmed $\Xi_{cc}$
baryons have two charm quarks and one light quark. On the other hand, 
$D$ meson has one charm quark and one light quark. The computed 
energy splittings between $\Xi_{cc}$ and the ground state of $D$ meson
are shown in \fgn{split_p_ccu_ccs_d_ds}. In the same figure
the energy splittings between $\Omega_{cc}$ and the ground
state of $D_s$ are shown. For each spin, the left column indicates splittings 
of $\Xi_{cc} - D$ while the right column shows $\Omega_{cc} - D_s$
splittings. The inset at the top figures shows the positive parity
ground states of spin-1/2 and spin-3/2 baryons. Results for $D$ and
$D_s$ mesons are taken from Ref.~\cite{Moir:2013ub}. By subtracting
the $D$ and $D_s$ meson masses from $\Xi_{cc}$ and $\Omega_{cc}$ baryons we
effectively leave only the energy of the excitation of a single charm quark in 
both cases. One would thus naively expect that both spectra will be
equivalent. This is almost true for the lowest state in each spin
parity channel, except for negative parity spin-7/2 state, as shown in
\fgn{split_p_ccu_ccs_d_ds}. That state is obtained from relativistic
operators and it is expected that such naive expectation may not hold
there. This is also seen for the excited states where the contribution
from relativistic operators is much larger. 
%We do not see
%this matching between $\Xi_{cc}$ and $\Omega_{cc}$ states.

As in our previous studies in charmonia~\cite{Liu:2012ze},
charmed-strange mesons~\cite{Moir:2013ub} and triply-charmed
baryons~\cite{Padmanath:2013zfa}, the systematic uncertainty
due to ${\cal{O}}(a)$ discretisation artefacts was investigated in another
calculation in which the spatial clover co-efficient was boosted from the
tree-level $c_s = 1.35$ to $c_s = 2$. At this value our extracted hyperfine
splitting was found to be physical~\cite{Liu:2012ze}. The mass
splittings were found to differ by around 40-45 MeV between the two
calculations. In this study also we found a similar mass difference between 
two calculations, indicating an equivalent scale for the uncertainties in this 
calculation.

%\clearpage
\bef[tbh]
\centering
\includegraphics[scale=0.35]{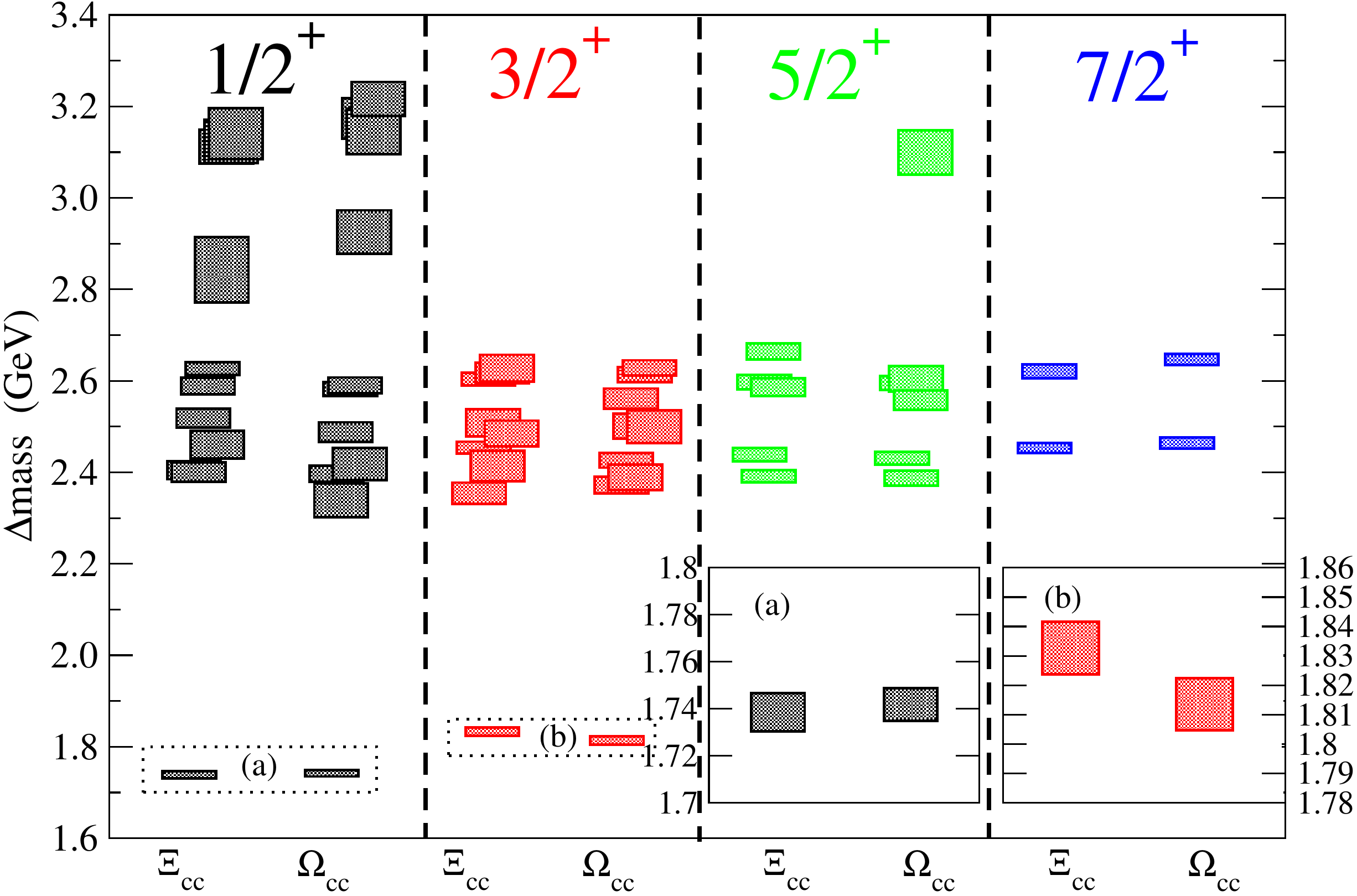}
%\vspace*{0.53in}
%\hspace*{-0.3in}
\includegraphics[scale=0.35]{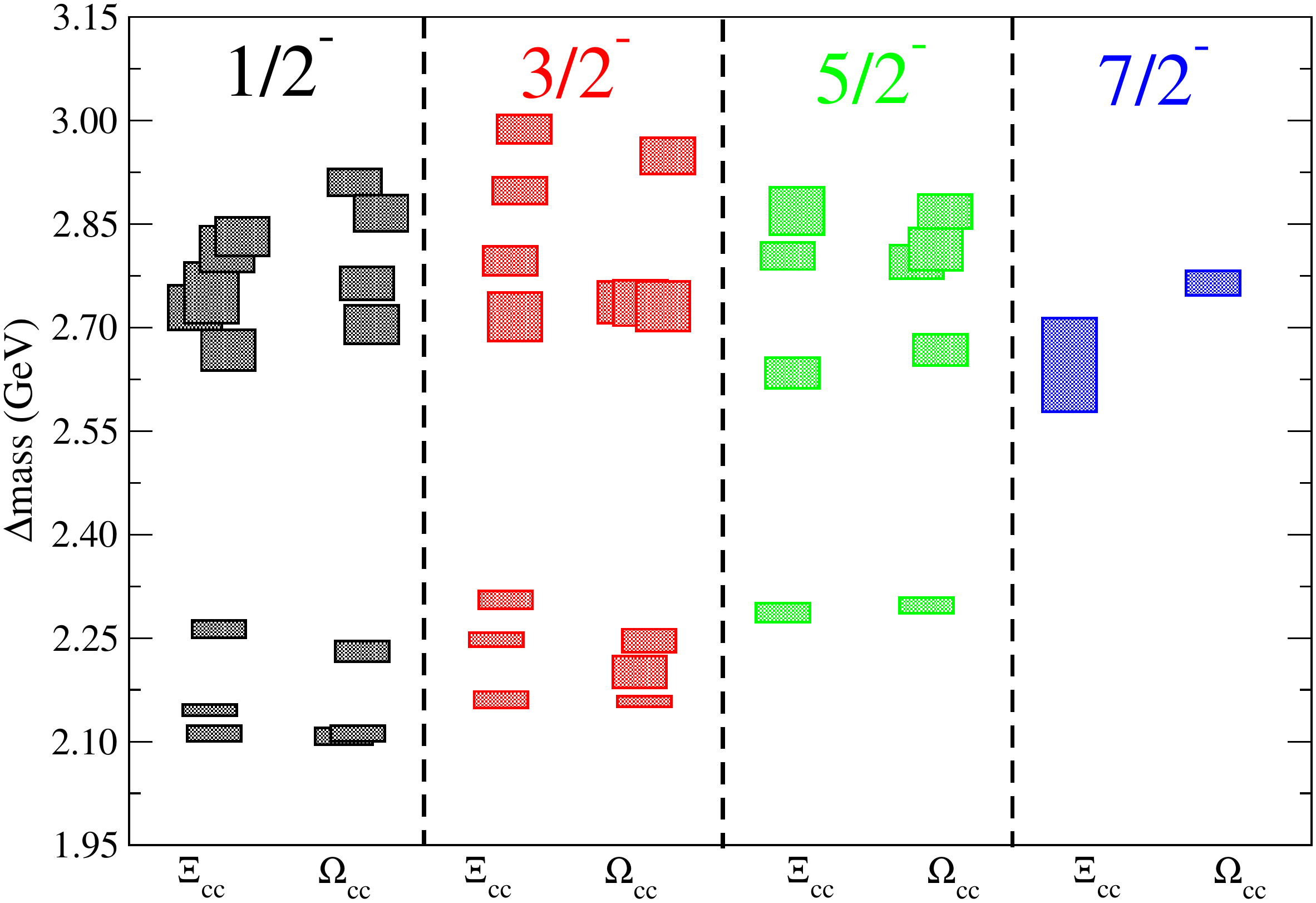}
\caption{Energy splittings of $\Xi_{cc}$ and $\Omega_{cc}$ baryons from $D$ and $D_s$ mesons, respectively. Splittings for the positive parity baryons are in the left plot and those for negative parity baryons are shown in the right plot. For each spin, left column is for splittings of  $\Xi_{cc}(ccu)$ baryons from the ground state $D(cu)$ meson and the right column is for splittings of  $\Omega_{cc}(ccs)$ baryons from the ground state $D_s(cs)$ meson. The inset at the top figures are for the positive parity ground states of spin-1/2 and spin-3/2 baryons. Results for $D$ and $D_s$ mesons are taken from Ref.~\cite{Moir:2013ub}.}
\vspace*{0.24in}
\eef{split_p_ccu_ccs_d_ds}

%%%%%%%%%%%%%%%%%%%%%%%%%%%%%%%%%%%%%%%%%%%%%%%%%
\subsection{Energy Splittings}
%%%%%%%%%%%%%%%%%%%%%%%%%%%%%%%%%%%%%%%%%%%%%%%%%
The energy splittings between various excitations in a spectrum provide
important information about the nature of interactions needed to
excite those states. The energy splittings also provide inputs for
building models to describe these states successfully. 
In \fgn{groundstate_split} we show energy splittings of the ground
states of each spin parity channel from the lowest state in that
parity channel.  For the positive parity, the lowest state is $J^{P} =
\frac12^{+}$ and for negative parity the lowest state is $J^{P} =
\frac12^{-}$.  It is interesting to note that both for $\Xi_{cc}$ and
$\Omega_{cc}$ this splittings are almost same. This indicates that the
interquark interactions, which are responsible for these splittings,
are similar in these two different type of baryons.

The most
notable spin dependent baryon energy splitting to consider is the hyperfine 
splittings between $\frac32^+$ and
$\frac12^+$ states, for example splitting between $\Delta$ and
nucleon. For doubly-charmed baryons we also compute this splitting and
show in \fgn{hfs_ccu_comp} for $\Xi_{cc}$ (left plot) and
$\Omega_{cc}$ (right plot) baryons. Our results (red circles) are
compared with other lattice results (blue squares) as well as with various model
results.  Note our results for $\Xi_{cc}$ are for 
pion mass 391 MeV. Results for ETMC \cite{Alexandrou:2012xk},
PACS-CS \cite{Namekawa:2013vu}, Bali {\it et. al}~\cite{Bali:2012ua},
and Briceno {\it et. al} \cite{Briceno:2012wt} are extrapolated to the
physical pion mass, while ILGTI \cite{Basak:2012py} results are at
pion mass 390 MeV.

In \fgn{hfs_p_comp} we compare these hyperfine splittings for
$\Xi_{cc}$ and $\Omega_{cc}$ baryons as a function of quark mass. In
the $x$-axis of that figure we use the square of the pseudoscalar
meson mass ($m_{ps}$) while $y$-axis shows hyperfine splittings at
those pseudoscalar meson masses. Along with $\Xi_{cc}$ and
$\Omega_{cc}$ we also show splittings between spin-$\frac32$ and
spin-$\frac12$ states of $\Omega_{ccc}$ baryon. The later spin dependent
splitting is not hyperfine in nature as both of them are of decuplet type. 
As in Ref.~\cite{Padmanath:2013zfa}, for the positive parity $\Omega_{ccc}$ baryons we take the spin-orbit splitting
between $E_{3}(\frac32^{+})$ and $E_{0}(\frac12^{+})$ states which have same
$L$ and $S$ values. For the negative parity $\Omega_{ccc}$ baryons we take
spin-orbit splitting between $E_{0}(\frac32^{-})$ and $E_{0}(\frac12^{-})$
states which also have same $L$ and $S$ values.

\bef[tbh] \centering
\vspace*{0.3in} \includegraphics[scale=0.32]{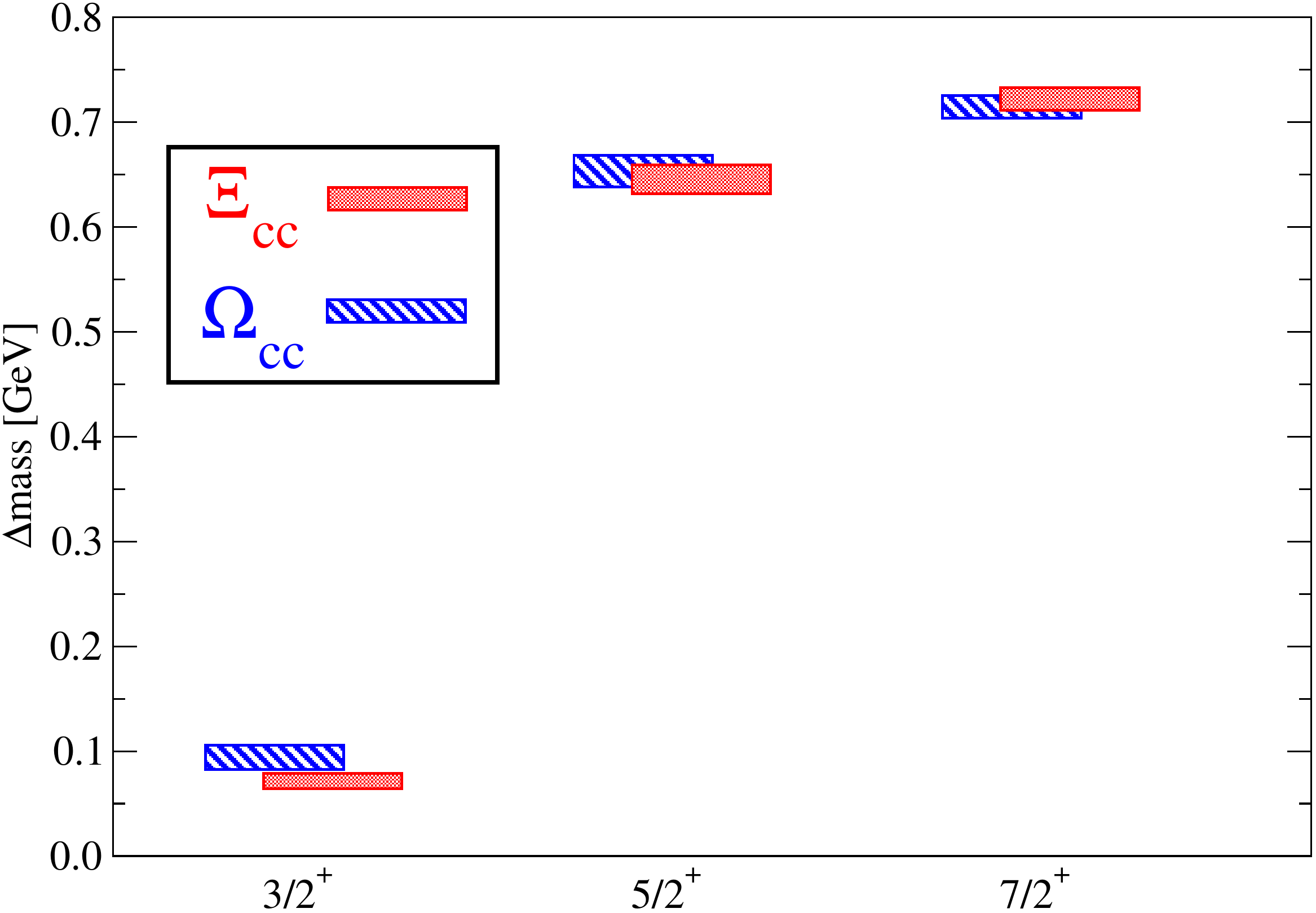}
\hspace*{0.1in} \includegraphics[scale=0.32]{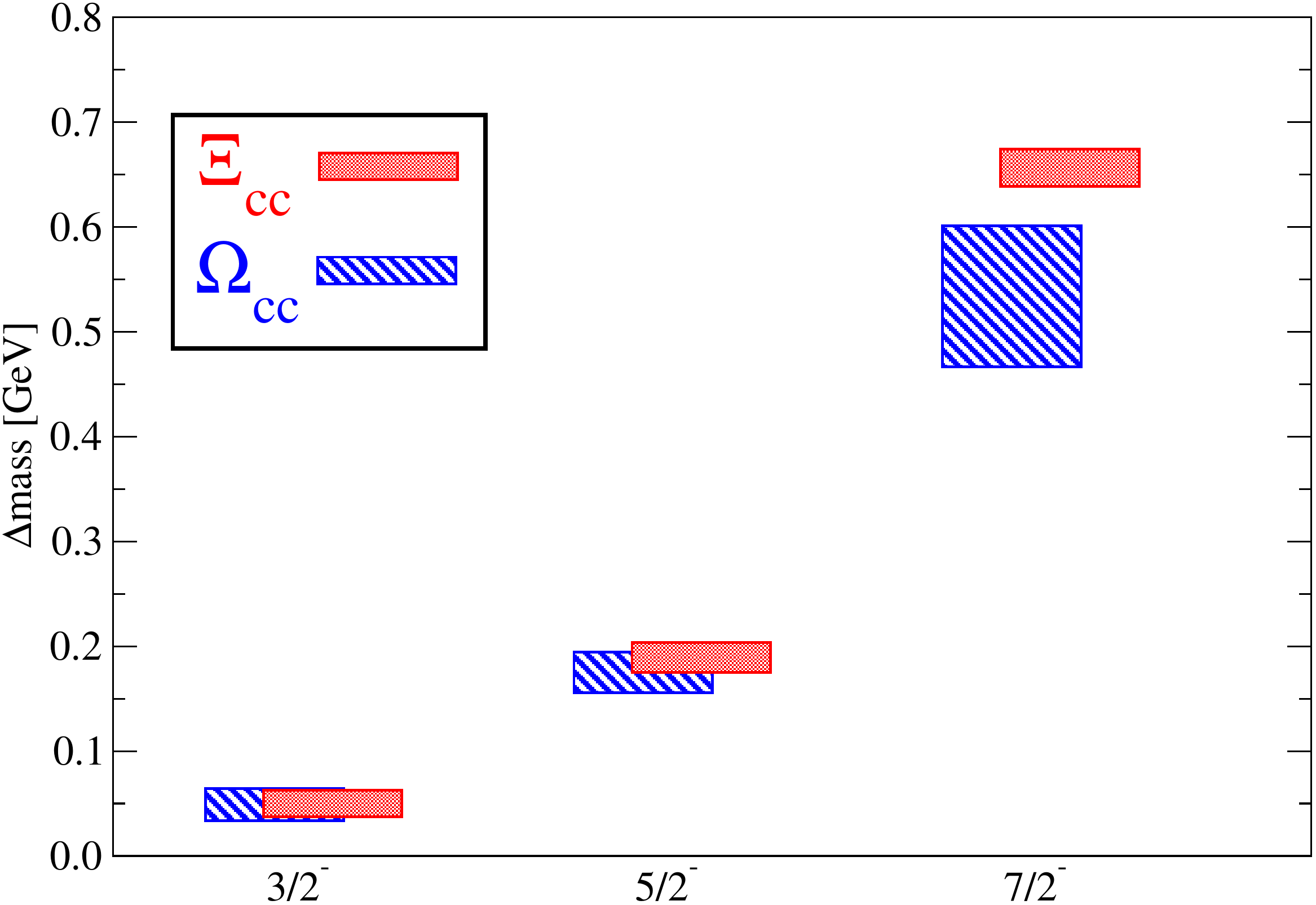}
%\vspace*{0.35in} \includegraphics[scale=0.32]{fig/ccu_split_3o2n.pdf}
%\hspace*{0.1in} \includegraphics[scale=0.32]{fig/ccs_split_3o2n.pdf}
\caption{Energy splittings (in units of GeV) of the ground states of
  each spin parity channel from the lowest state in that parity
  channel. Splittings for $\Xi_{cc}$ are shown with red color
  rectangle and those of $\Omega_{cc}$ are represented by blue color
  shredded rectangle.  For the positive parity, the lowest state has
  $J^{P} = \frac12^{+}$ and for the negative parity the lowest state
  has $J^{P} = \frac12^{-}$.}
\hspace*{0.1in} \eef{groundstate_split}

%newpage
%fgn{hfs_ccs_comp}

\bef[tbh]
\centering
\includegraphics[width=8.5cm,height=7cm]{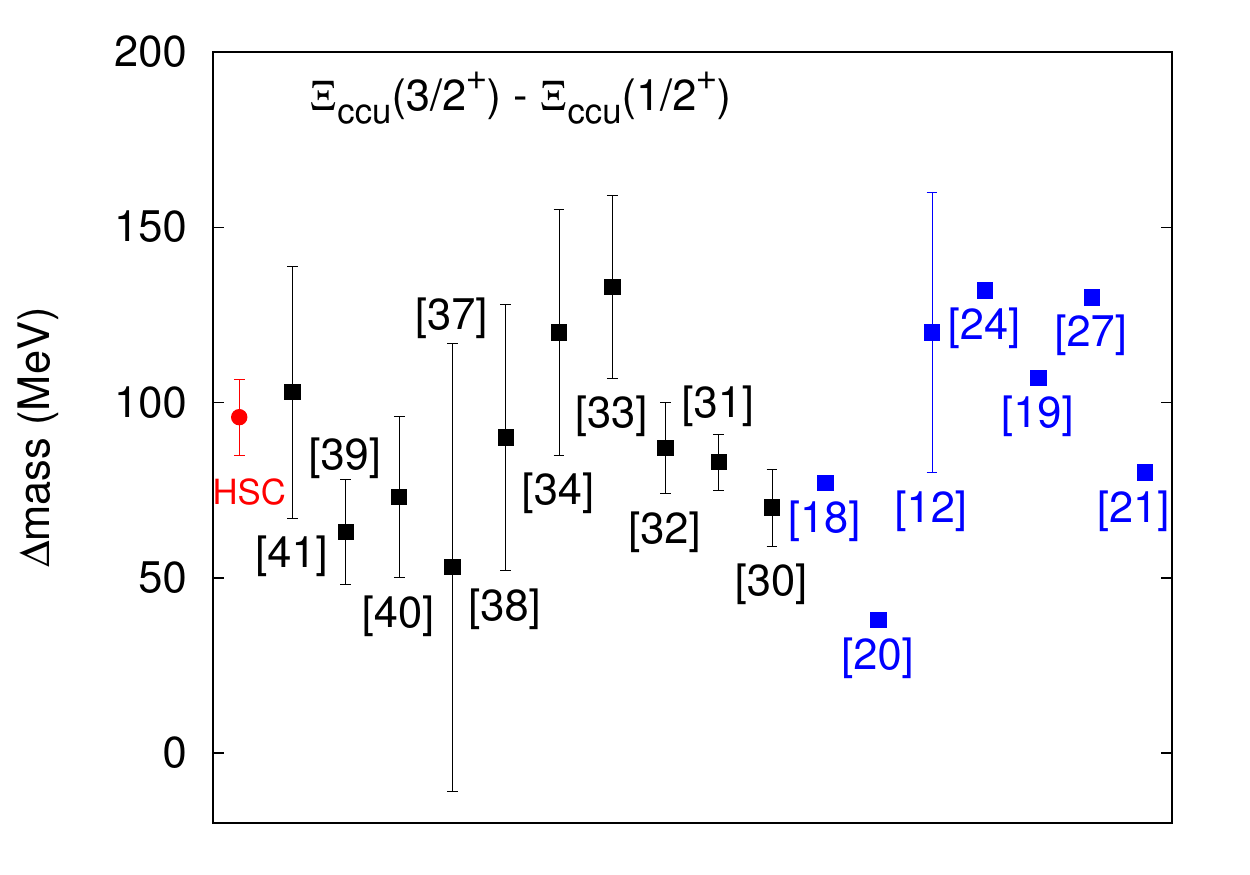}
\includegraphics[width=8.5cm,height=7cm]{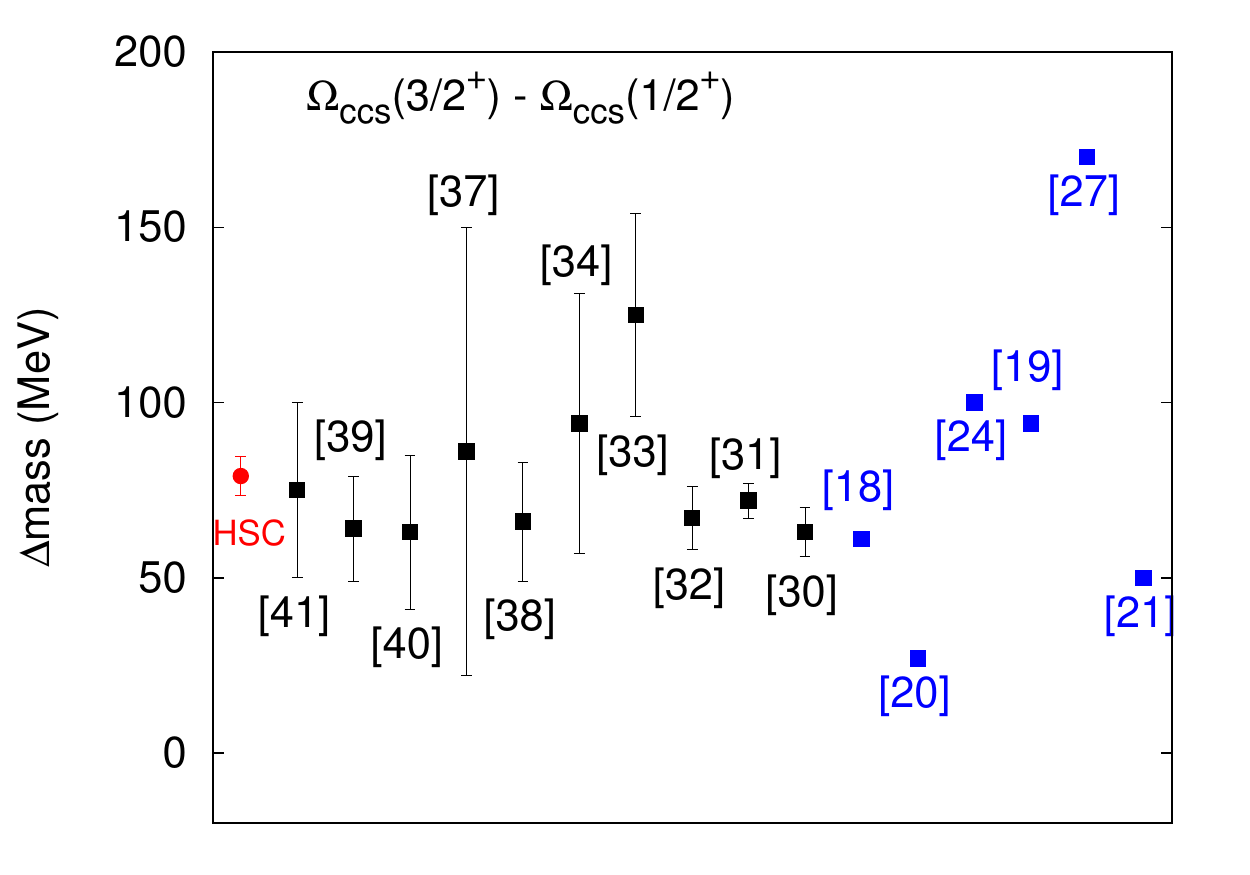}
\caption{Hyperfine mass splittings between spin-$\frac32^{+}$ and spin-$\frac12^{+}$ states of $\Xi_{cc}$ (left plot) and $\Omega_{cc}$ (right plot) baryons are compared for various lattice and model results. Our results are shown by red filled circle (HSC). Results from this work are at pion mass 392 MeV,
  results for ETMC \cite{Alexandrou:2012xk}, PACS-CS
  \cite{Namekawa:2013vu}, Bali {\it et. al}~\cite{Bali:2012ua}, and
  Briceno {\it et. al} \cite{Briceno:2012wt} are extrapolated to the
  physical pion mass, while ILGTI \cite{Basak:2012py} results are at
  pion mass 390 MeV and 340 MeV respectively.}
%\vspace*{0.42in}
\eef{hfs_ccu_comp}

\bef[tbh] \centering
\includegraphics[scale=0.7]{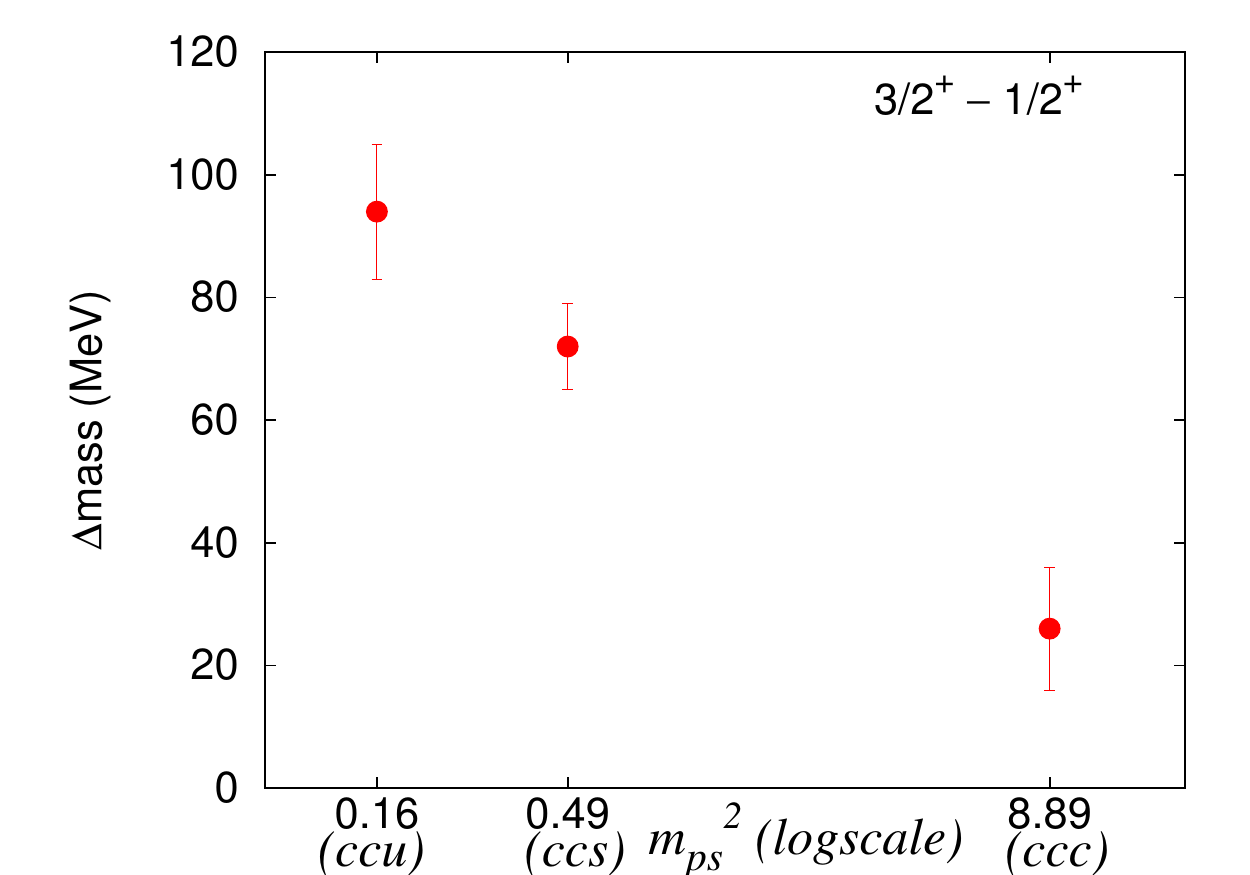}
\includegraphics[scale=0.7]{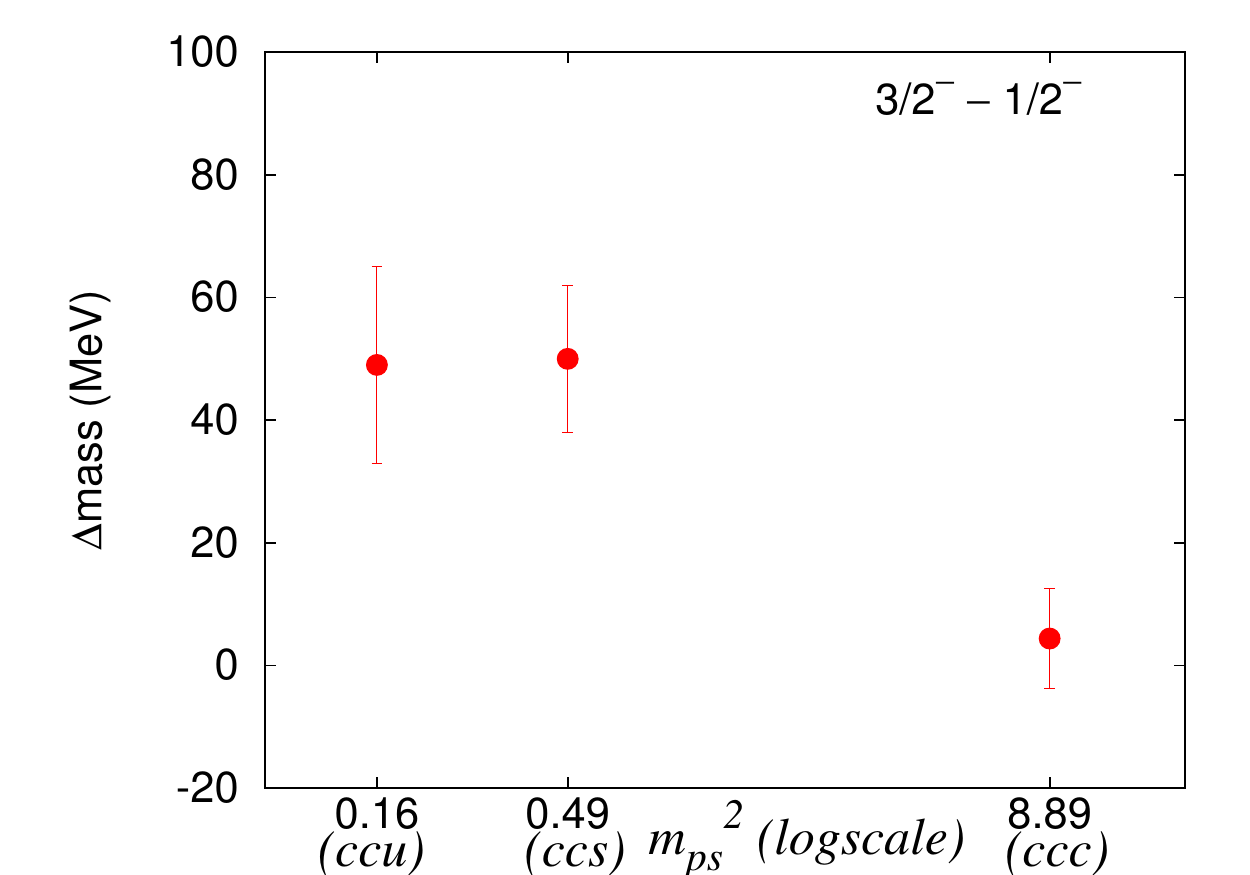}
\caption{Mass splittings between spin-$\frac32^{+}$ and
  spin-$\frac12^{+}$ states (left figure) and between
  spin-$\frac32^{-}$ and spin-$\frac12^{-}$ ground states (right
  figure) are compared for $\Xi_{cc}(ccu)$, $\Omega_{cc}(ccs)$ and
  $\Omega_{ccc}(ccc)$ baryons. For $\Omega_{ccc}$, mass splitting is
  between $E_{0}(1/2^{+})$ and $E_{3}(3/2^{+})$ states, which is
  actually due to spin-orbit coupling, while for $\Xi_{cc}$ and
  $\Omega_{cc}$, these are between respective ground states and are
  due to hyperfine splittings.}
\vspace*{0.2in} \eef{hfs_p_comp}

In Ref.~\cite{Padmanath:2013zfa} we evaluated energy splittings
between triply-flavoured baryons{$^{1}${\footnotetext[1]{by
      triply-flavoured baryons, we mean $\Delta\,,
      \Omega\,,\Omega_{ccc}$ and $\Omega_{bbb}$ baryons.}}} and
studied their quark mass dependence. It was observed that various
splittings decrease significantly with quark masses.  Following heavy
quark effective theory one can expand the mass of a heavy hadron, with
$n$ heavy quarks, as $M_{H_{nq}} = nM_Q + A + B/m_Q +
\mathcal{O}(1/m_{Q^2})$~\cite{Jenkins:1996de}. With that expectation,
we fitted various energy splittings of triply-flavoured baryons with a
form $a + b/m_{ps}$, with $m_{ps}$ the pseudoscalar meson mass, and obtained 
good fits. Motivated by that, here we also fit similar energy splittings
for doubly-charmed baryons. There is a however a difference
for doubly-charmed baryons because to show quark mass dependence one
needs to use triply-charmed baryons data also.  For the triply-flavoured
baryons in Ref.~\cite{Padmanath:2013zfa}, states at different quark
masses are obtained from the same operators and so are simple to
compare. However, in this work, doubly- and triply-charmed baryons are not 
obtained from the same operators except for positive parity spin 3/2 and 
spin-7/2 states.  We thus consider energy
splittings only for these states. More specifically, we calculate the
following energy splittings : $\Xi_{cc}^{*}(ccu) - D_{u}(\bar{c}u)\,,
\Omega_{cc}^{*}(ccs)- D_{s}(\bar{c}s)$ and $\Omega_{ccc}^{*}(ccc)-
\eta_{c}(\bar{c}c)$, and $ \Xi_{cc}^{*}(ccu) - D^{*}_{u}(\bar{c}u)\,,
\Omega_{cc}^{*}(ccs)- D^{*}_{s}(\bar{c}s)$ and $\Omega_{ccc}^{*}(ccc)-
J/\psi(\bar{c}c)$ and plot in \fgn{energy_split32}.  As in
Ref.~\cite{Padmanath:2013zfa}, we fit these splittings to the 
form $a + b/m_{ps}$ and obtain reliable agreement. Energy splittings from
vector mesons are also fitted with a constant. We emphasis that energy 
splittings at very light quark masses should not be calculated using the above 
form, which is valid only for heavy quarks.

It is interesting to note that one can extrapolate the fitted results
to the bottom mass to obtain the energy splittings of
$\Omega^{*}_{ccb}(\frac32^+) - B_{c}$ and $\Omega^{*}_{ccb}(\frac32^+)
- B_{c}^{*}$ at the mass of $B_c$ meson.  The difference between these 
two quantities yields  the
mass splitting of $B_c^{*} - B_c$. Using this, we obtain the mass
splitting of $B_c^{*} - B_c$ as $80 \pm 8$ MeV (fitting with a form $a +
b/m_{ps}$) and $76 \pm 7$ MeV (fitting with a constant term). This result
agrees very well with those obtained by potential
models~\cite{Eichten:1994gt}. It does not however agree with
other lattice QCD prediction~\cite{Gregory:2009hq, Gregory:2010gm,
  McNeile:2012qf} which provide much lower values. It will be
interesting to study the effects of the lattice cut-off on the hyperfine
splittings of these hadrons with the possibility of future discovery of 
$B^{*}_c$ meson. Only one charmed-bottom meson has been discovered so 
far~\cite{PDG}. Assuming it as a
pseudoscalar meson with mass 6277 MeV and using the extrapolated value of $\Omega^{*}_{ccb}(\frac32^+) - B_{c}$, we empirically predict the
mass of $\Omega_{ccb}^{*}(3/2^+)$ to be 8050 $\pm$ 10 MeV which is
consistent with the results from various models
\cite{Hasenfratz:1980ka, Martynenko:2007je, SilvestreBrac:1996bg} and a recent lattice calculation~\cite{Brown:2012qx}.

\bef[tbh]
\centering
\vspace*{-0.1in}
\includegraphics[scale=0.7]{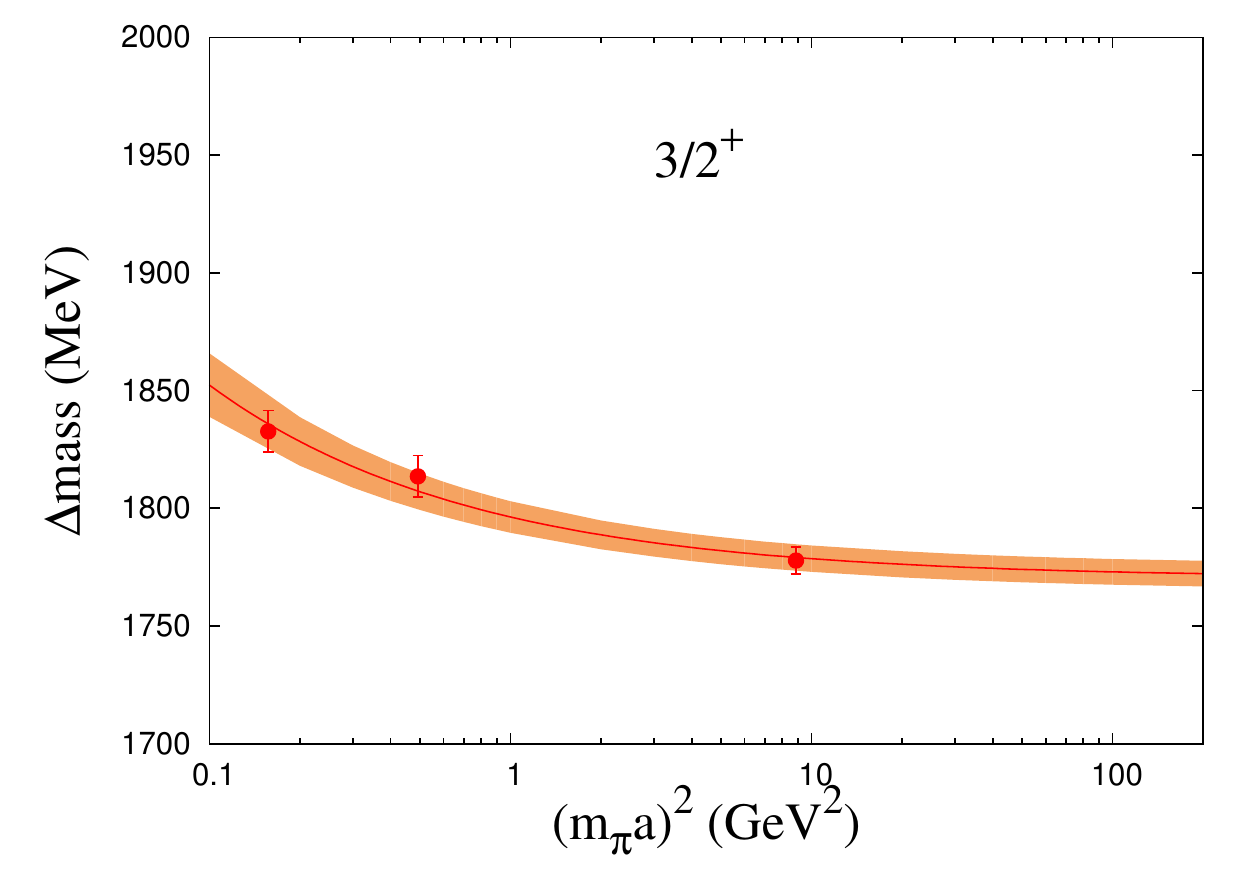}
\includegraphics[scale=0.7]{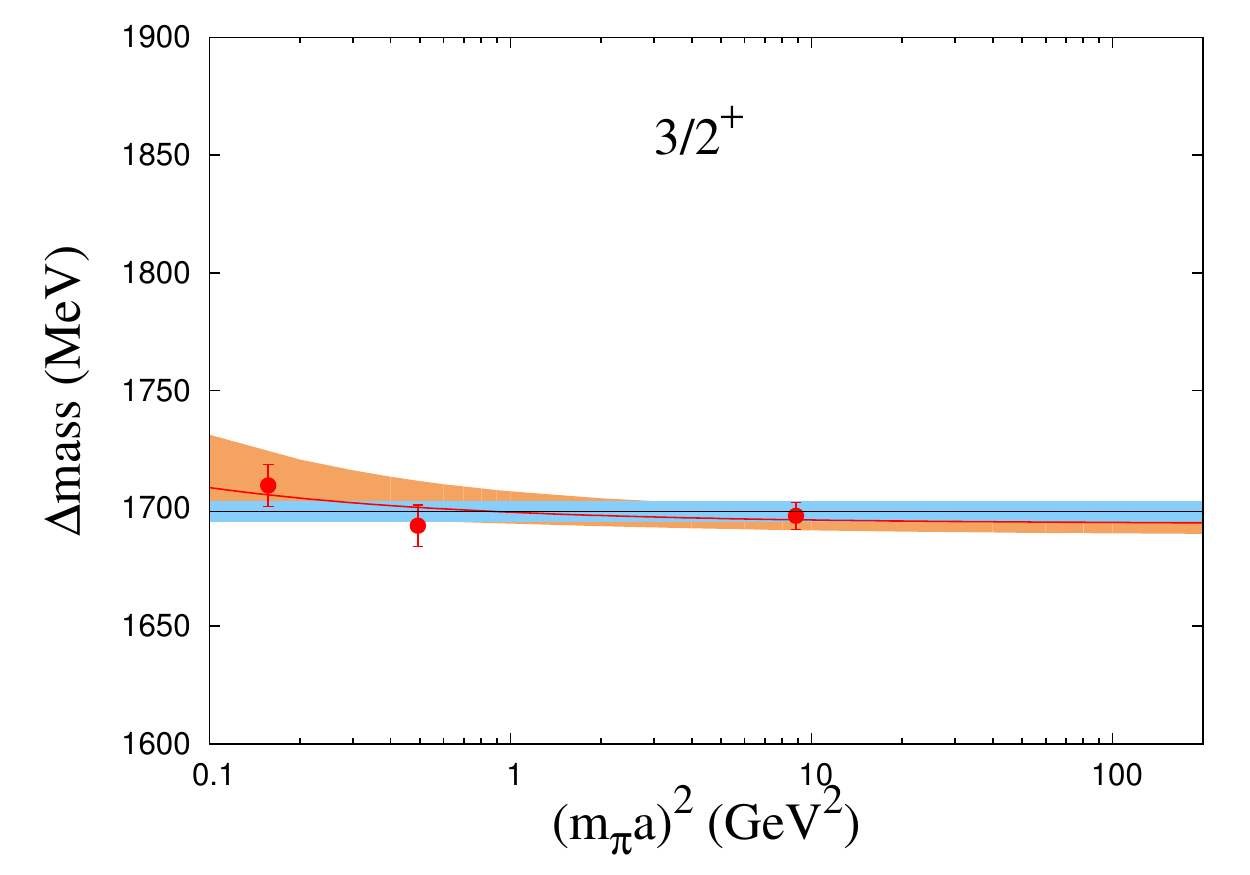}
\caption{Energy splittings between positive parity spin-3/2 baryons and pseudoscalar meson as well as vector mesons are plotted against the square of the pseudoscalar masses. Left figure is the splittings of  : $\Xi_{cc}^{*}(ccu) - D_{u}(\bar{c}u)\,, \Omega_{cc}^{*}(ccs)- D_{s}(\bar{c}s)$ and $\Omega_{ccc}^{*}(ccc)- \eta_{c}(\bar{c}c)$, and right figure includes following splittings : $ \Xi_{cc}^{*}(ccu) - D^{*}_{u}(\bar{c}u)\,, \Omega_{cc}^{*}(ccs)- D^{*}_{s}(\bar{c}s)$ and $\Omega_{ccc}^{*}(ccc)- J/\psi(\bar{c}c)$; they are plotted against the square of the pseudoscalar masses (i.e., at $D_{u}\,, D_{s}$) and $\eta_c$ mass.  We fit the quark mass dependence 
with a form $a + b/m_{ps}$ (right figure is also fitted with a constant term). The fitted results are shown by solid lines with shaded regions as one sigma errorbars.}
%\vspace*{0.32in}
\eef{energy_split32}

%\fgn{hfs_ccs_comp}
%\fgn{hfs_ccs_comp}
%==================================================================
%==================================================================
%==================================================================
%==================================================================
\section{Conclusions}
In this work, results from the first non-perturbative calculation on
the excited state spectroscopy of the doubly-charmed baryons with spin
up to 7/2 are presented. We performed our calculations using
lattice QCD with dynamical $2+1$ flavours clover quarks on anisotropic 
lattices. There is no clear experimental observation to date of a  
doubly-charmed baryon. It is important for theory to determine if various 
model results using quark-diquark symmetry or otherwise, can be tested with
a non-perturbative method for these systems for which the interaction is
governed by two widely separated scales. Our extensive study of doubly
charmed baryons has made a first attempt to connect possible experimental and
theoretical studies.

We employ a large basis of creation operators and use the distillation
technique to calculate the doubly-charmed baryon correlation
functions. A variational fitting method was employed to extract the
spectrum. We observe approximate rotational symmetry for these
operators at the scale of hadrons. Having realized a good rotational
symmetry we determine the overlap factors, as in previous
studies~\cite{Dudek:2007wv, Dudek:2010wm, Dudek:2009qf, Dudek:2010ew,
  Edwards:2011jj,Dudek:2012ag, Edwards:2012fx,
  Liu:2012ze,Moir:2013ub,Padmanath:2013zfa}. We are able to extract
states reliably with spin up to 7/2 and also studied the mixing
between various operators.

The main results are shown in \fgn{split_ccu} and \fgn{split_ccs} for
$\Xi_{cc}$ and $\Omega_{cc}$ baryons, respectively. The ground states
for spin 1/2 and 3/2, which have more relevance to recent experiments,
are shown in \fgn{groundstate_ccs} and \fgn{groundstate_ccu}.  As in
Ref.~\cite{Edwards:2012fx,Padmanath:2013zfa}, we also find bands of
states with alternating parities and increasing energies. We also
observed the number of extracted states of each spin in the three
lowest-energy bands and the number of quantum numbers expected based
on weakly broken $SU(6)\times O(3)$ symmetry agree perfectly, {\it
  i.e.,} the doubly-charmed baryon spectra remarkably resemble the
expectations of quantum numbers from non-relativistic quark
model~\cite{Greenberg:1964pe,Isgur:1977ef,Isgur:1978xj}.  This
symmetry was also observed for light, strange~\cite{Edwards:2012fx}
and triply-charmed baryon spectra~\cite{Padmanath:2013zfa}.  For
positive parity states this agreement does not get spoiled even with
the inclusion of non-relativistic hybrid operators. However, it is
expected that this band structure will persist with the
inclusion of relativistic operators which contribute more to the
higher excited states. The extracted spectra for the higher excited
states also support this observation.  Note our operator set does not include 
any multi-hadron operators. It is
expected that inclusion of those operators, particularly those
involving light quarks, may modify some of these conclusions.  We
are also able to decode the structure of operators leading to a
particular state: whether constructed by relativistic,
non-relativistic, hybrids, non-hybrid types or a mixture of them
all. However, this identification is not possible for negative parity
states and highly excited positive parity states, as argued in
Ref.~\cite{Edwards:2012fx}.

Our calculated spectra do not support the chiral multiplet structure
of doubly heavy baryons speculated in
Refs. ~\cite{Bardeen:2003kt,Quigg:2011sb}. The ground states of doubly
charmed baryons are not degenerate and both for $\Xi_{cc}$ and
$\Omega_{cc}$ hyperfine splittings are around $80-100$ MeV, as shown
in fig 15. The calculated spectra do not match the diquark picture 
and instead they are remarkably similar to the expectations from
non-relativistic quark models with an SU(6)$\times$ O(3) symmetry.

The study of the energy splittings between various excited states is  
quite helpful in revealing the nature of interquark interactions and
a detailed knowledge of them could help to build successful models. We
calculated hyperfine mass splittings between spin-$\frac32^{+}$ and
spin-$\frac12^{+}$ states of $\Xi_{cc}$ and $\Omega_{cc}$ and compared
those with various other lattice and model predictions.

Energy splittings between the ground states of different spins were
also evaluated and we observed that the hierarchy of the
first few energy excitations in $\Xi_{cc}$ and $\Omega_{cc}$ are quite
similar. This indicates the involvement of similar dynamics to excite
these states. To study the quark mass dependence of the energy splittings
we compared results for $\Xi_{cc}(ccu)$, $\Omega_{cc}(ccs)$ and
$\Omega_{ccc}(ccc)$ baryons for which there is a common `cc' diquark
and a varying quark from light to charm. %`u' to `c'. 
Encouraged by a successful fitting
of the mass splittings in triple-flavoured
baryons~\cite{Padmanath:2013zfa} we studied similar mass splittings of
doubly-charmed baryons. Here also we find that a heavy quark motivated
form $a+b/m_{ps}$ can fit quite successfully energy splittings like :
$\Xi_{cc}^{*}(ccu) - D_{u}(\bar{c}u)\,, \Omega_{cc}^{*}(ccs)-
D_{s}(\bar{c}s)$ and $\Omega_{ccc}^{*}(ccc)- \eta_{c}(\bar{c}c)$, and
$ \Xi_{cc}^{*}(ccu) - D^{*}_{u}(\bar{c}u)\,, \Omega_{cc}^{*}(ccs)-
D^{*}_{s}(\bar{c}s)$ and $\Omega_{ccc}^{*}(ccc)-
J/\psi(\bar{c}c)$. From the fitted results we are able to predict
$B_c^{*} - B_c$ = $80 \pm 8$ MeV and $\Omega_{ccb}^{*}(3/2^+)$ = 8050
$\pm$ 10 MeV.

%==================================================================
%==================================================================
%==================================================================
%==================================================================
\section{Acknowledgements}
We thank our colleagues within the Hadron Spectrum Collaboration.
Chroma [43] and QUDA [44, 45] were used to perform this work on the
Gaggle and Brood clusters of the Department of Theoretical Physics,
Tata Institute of Fundamental Research, at Lonsdale cluster maintained by
the Trinity Centre for High Performance Computing funded through
grants from Science Foundation Ireland (SFI), at the SFI/HEA Irish
Centre for High-End Computing (ICHEC), and at Jefferson Laboratory
under the USQCD Initiative and the LQCD ARRA project.  Gauge
configurations were generated using resources awarded from the
U.S. Department of Energy INCITE program at the Oak Ridge Leadership
Computing Facility at Oak Ridge National Laboratory, the NSF Teragrid
at the Texas Advanced Computer Center and the Pittsburgh Supercomputer
Center, as well as at Jefferson Lab. MP acknowledges support from the Trinity College Dublin Indian Research Collaboration Initiative,
   Graduate school Tata Institute of Fundamental Research Mumbai and Austrian Science Fund
   (FWF):[I1313-N27]; NM acknowledges support from Department of Theoretical Physics, TIFR;
RGE acknowledges support from U.S. Department of Energy
contract DE-AC05-06OR23177, under which Jefferson Science Associates,
LLC, manages and operates Jefferson Laboratory; 
%==================================================================
%==================================================================

\end{document}